\documentclass[preprint,aps,prd,showpacs,superscriptaddress,nofootinbib]{revtex4-1}
\usepackage{amsmath,amssymb,amsbsy,amstext,amsthm,simplewick,amsfonts} 
\usepackage{physics}
\usepackage{graphicx} 
\usepackage{dcolumn} 
\usepackage{bm} 
\usepackage{comment}
\usepackage{enumitem}
\usepackage{hyperref}
\usepackage{tablefootnote}
\usepackage{slashed}
\usepackage{multirow}

\renewcommand{\vec}[1]{\mathbf{#1}}

\hypersetup{
	colorlinks = true,
	linkcolor = blue,
	citecolor = blue,
	urlcolor = blue
}

\begin{document}

\title{
Displaced Signals from Long-lived Particles\\in Neutrinoless Double Beta Decay
}

\author{Patrick D. Bolton}
\email{patrick.bolton@ijs.si}
\affiliation{Jožef Stefan Institute, Jamova 39, 1000 Ljubljana, Slovenia}

\author{Noor-In\`es Boudjema}
\email{noor-ines.boudjema.19@ucl.ac.uk}
\affiliation{University College London, Gower Street, London WC1E 6BT, United Kingdom}

\author{Frank F. Deppisch}
\email{f.deppisch@ucl.ac.uk}
\affiliation{University College London, Gower Street, London WC1E 6BT, United Kingdom}

\author{Chandan Hati}
\email{chandan@ific.uv.es}
\affiliation{Instituto de F\'isica Corpuscular (CSIC-Universitat de Val\`encia), 46980 Paterna, Spain}

\author{Chayan Majumdar}
\email{chayanmajumdar@impcas.ac.cn}
\affiliation{Institute of Modern Physics, Chinese Academy of Sciences (IMP, CAS), Lanzhou 730000, Gansu, China}

\begin{abstract}
Much of the literature on neutrinoless double beta ($0\nu\beta\beta$) decay with light new-physics mediators focuses on invisible missing energy. We investigate the scenario in which a massive Majoron-like particle $\phi$ is produced on-shell during $0\nu\beta\beta$ decay and subsequently decays into visible final states after travelling a macroscopic distance. Specifically, we analyze the displaced energy deposition from $\phi$ decays into a photon pair ($\gamma\gamma$), a photon and a dark photon ($\gamma\gamma_D$) and an electron-positron pair ($e^+e^-$). We demonstrate that the displaced decays modify the expected visible energy spectra and provide novel, distinct experimental signatures at current and upcoming $0\nu\beta\beta$ experiments, with the promise of improving the sensitivity of the standard invisible Majoron searches in $0\nu\beta\beta$ decay by more than two orders of magnitude. The relevant effective couplings can naturally arise in well-motivated ultraviolet-complete scenarios that conventional $0\nu\beta\beta$ decay searches cannot probe.
\end{abstract}

\maketitle
\raggedbottom
\section{Introduction}
\label{sec:intro}

Searches for nuclear double beta decays and related processes play a critical role in our understanding of neutrino physics and potential physics beyond the Standard Model (SM) in general. While the SM permits two-neutrino double beta ($2\nu\beta\beta$) decay~\cite{Goeppert-Mayer:1935uil}, an ultra-rare process with half-lives exceeding $T_{1/2}^{2\nu} \approx 10^{19}$~yrs, it provides a high-statistics baseline for modern experiments~\cite{Barabash:2019nnr}. The primary experimental thrust remains the search for neutrinoless double beta ($0\nu\beta\beta$) decay~\cite{Furry:1939qr}, which would signal lepton number violation and confirm the Majorana nature of neutrinos~\cite{Majorana:1937vz, Deppisch:2012nb, Graf:2018ozy, Cirigliano:2018yza, Deppisch:2020ztt}.

As experimental exposures and sensitivities increase, $2\nu\beta\beta$ data are increasingly utilized to search for subtle signatures of new physics~\cite{Simkovic:2018rdz}. Deviations from the expected spectral shape can indicate non-standard neutrino interactions or the emission of neutral exotic particles~\cite{Deppisch:2020mxv, Deppisch:2020sqh, Bolton:2020ncv, Agostini:2020cpz, Cepedello:2018zvr}. While much of the existing literature focuses on missing energy, the emission of extra visible particles, namely electrons, positrons, or photons, is rarely studied. For example, a search for neutrinoless triple beta ($0\nu 3\beta$) decay was discussed in \cite{Barabash:2019enn} as a test of Lorentz invariance. Neutrinoless quadruple beta ($0\nu 4\beta$) decay \cite{Heeck:2013rpa} can arise if lepton number is broken by four units but it is strongly suppressed and effectively ruled out by the non-observation of an exotic contribution to $2\nu\beta\beta$ decay~\cite{Deppisch:2020sqh}.

The most prominent candidate for the emission of an exotic particle is the Majoron, originally conceived as a massless Goldstone boson~\cite{Chikashige:1980ui, Gelmini:1980re}. However, it can be generalised to Majoron-like models to include massive scalars or vectors that may couple to both neutrinos and other light sectors~\cite{Burgess:1992dt, Burgess:1993xh, Carone:1993jv, Bamert:1994hb, Hirsch:1995in, Blum:2018ljv}. If such a particle is lighter than the $Q$-value of the $0\nu\beta\beta$ isotope, it can be produced on-shell. Usually, such a massive Majoron-like particle is considered to be emitted as a real particle and its decay is disregarded. This is appropriate as long as the decay width is small and the available final states are all invisible. Current limits~\cite{Boudjema:2025okq, PandaX:2025tls} on the coupling to neutrinos are such that the Majoron-like particle can have macroscopic proper decay lengths, $L_\phi \approx 0.1~\text{cm} \times (10^{-4} / c_\nu)^2 (1~\text{MeV} / m_\phi)$. Combined with the ton- and multiton-scale dimensions of current and future double beta decay experiments, this motivates the search for long-lived MeV-scale exotic particles.

\begin{figure}[t!]
    \centering
    \includegraphics[width=0.99\textwidth]{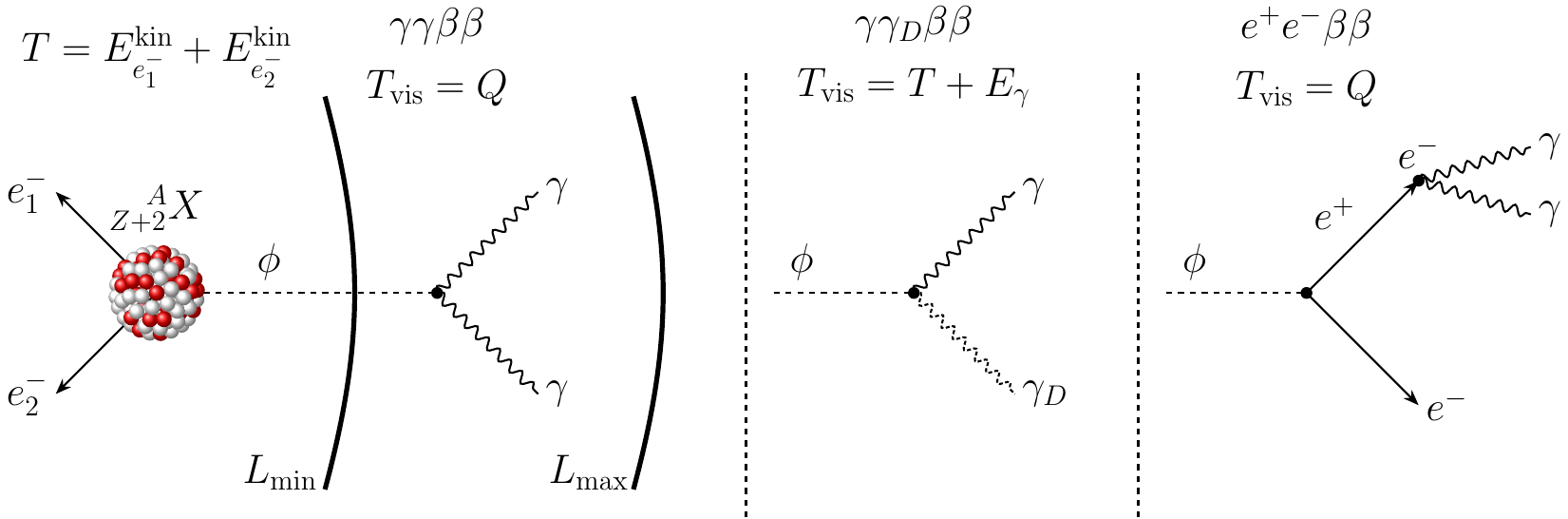}
    \caption{Neutrinoless double beta decay associated with the emission of a long-lived, light pseudoscalar $\phi$ that decays over the macroscopic distance $L$ to two photons ($\gamma\gamma\beta\beta$), a photon and a dark photon ($\gamma\gamma_D\beta\beta$), and an electron-positron pair ($e^+e^-\beta\beta$). The prompt electron kinetic energy release $T$ and the displaced energy release add up to the visible energy $T_\text{vis}$ as indicated.}
\label{fig:diagram}
\end{figure}

In this paper, we investigate a scenario in which a massive Majoron-like scalar is produced on-shell and subsequently decays into visible final states after traveling a potentially macroscopic distance; see Fig.~\ref{fig:diagram}. While some previous studies have examined portals to exotic fermions that could serve as dark matter~\cite{Huang:2014bva, Nozzoli:2022tov} with an invisible final-state signature, we focus here on displaced charged or radiative final state signatures of this mediator. Specifically, we analyze the production of photons, possibly in association with dark photons, and $e^+ e^-$ pairs arising from the pseudoscalar decay, which would produce displaced and delayed signatures inside double beta decay detectors, modify the double beta decay spectrum and therefore provide distinct and novel experimental signatures.

Current experiments such as KamLAND-Zen \cite{KamLAND-Zen:2019imh} and NEMO-3 \cite{NEMO-3:2009fxe} provide the precision necessary to resolve these spectral modifications. We expect that displaced vertices of this kind are resolvable across several double beta decay technologies. Segmented bolometer arrays such as CUORE \cite{CUORE:2024fak, CUORE:2024ikf} can identify a secondary vertex through crystal multiplicity, while liquid-scintillator detectors namely, KamLAND-Zen \cite{KamLAND-Zen:2019imh}, SNO+ \cite{SNO:2021xpa}, JUNO \cite{Zhao:2016brs, Liu:2018fpq} and time projections chambers like PandaX-III \cite{Zhang:2023ywy} continuously reconstruct the event position, allowing a secondary vertex to be identified directly. The laboratory-based searches will provide an excellent complement to the existing astrophysical and cosmological bounds, and searches by other facilities for light new physics~\cite{KamLAND-Zen:2012uen, Kharusi:2021jez, CUPID-0:2022yws, Dev:2025tdv}. In the following, we will show that displaced modes of $0\nu\beta\beta$ decay accompanied by di-photon ($\gamma\gamma\beta\beta$), photon-dark photon ($\gamma\gamma_D\beta\beta$), and electron-positron ($e^+e^- \beta\beta$) pair final states open up the possibility of searching for long-lived particles in double beta decay experiments. 

The paper is organized as follows. In Sec.~\ref{sec:model}, we introduce the effective interactions and decay modes of $\phi$. In Sec.~\ref{sec:constraints}, we summarize constraints on our scenario from other contexts, while the signatures in double beta decay experiments are analyzed in Sec.~\ref{sec:dbd}. Finally, we discuss the sensitivity of current and future experiments in Sec.~\ref{sec:sensitivity}, along with an outlook.

%

\section{Pseudoscalar coupling to fermions and photons}
\label{sec:model}

We are interested in a scenario where an additional light on-shell Majoron-like particle $\phi$ is emitted in the final state of the usual $0\nu\beta\beta$ decay. Subsequently, $\phi$, henceforth assumed to be an arbitrary light pseudoscalar particle for concreteness, decays (partly) into visible SM final states (such as photons or charged leptons) within the detector after traveling a finite distance from the primary $0\nu\beta\beta\phi$ decay vertex. In general, $\phi$ can be emitted from a contact or long-range realization of the $0\nu\beta\beta\phi$ decay, leading to a number of possibilities for the UV origin of interactions leading to $\phi$. Here, we will focus on a minimal scenario where $\phi$ has effective interactions with the SM neutrinos and charged leptons,
\begin{align}
\label{eq:lepton_couplings}
    \mathcal{L} 
    = \frac{1}{2} c_\nu^{\alpha\beta} (\bar\nu_\alpha P_L \nu_\beta) \phi 
    + c_e^{\alpha\beta}(\bar{e}_\alpha P_L e_\beta) \phi + \text{h.c.},
\end{align}
where $\alpha,\beta = e,\mu,\tau$ correspond to the three SM flavours. Note that even if such interactions are absent at tree-level, they will necessarily be induced at loop level via mixing of $\phi$ and SM $Z$ via self-energy type of diagrams involving the neutrinos. For most of our discussion here, we will be interested in the $\phi$ couplings to the electron flavor, i.e., $c_\nu^{ij} =  i U_{ei} U_{ej} c_\nu$ and $c_e^{ij} = i\delta_{ei}\delta_{ej} c_e$ in the mass basis, where $c_\nu$ and $c_e$ are real parameters and $U$ is the Pontecorvo–Maki–Nakagawa–Sakata (PMNS) matrix. To facilitate interesting vector final states in $\phi$ decay, including photons, we will also consider the effective vector portals 
\begin{align}
\label{eq:gauge_couplings}
    \mathcal{L} = 
      \frac{g_{\phi\gamma\gamma}}{4}   \phi F_{\mu\nu}\tilde{F}^{\mu\nu} 
    + \frac{g_{\phi\gamma\gamma_D}}{2} \phi F_{\mu\nu}\tilde{F}_D^{\mu\nu},
\end{align}
where $F_{\mu\nu}$ and ${F_D}_{\mu\nu}$ are, respectively, the field strength tensors of the photon and a dark photon corresponding to a dark Abelian $U(1)_D$ extension of the SM. The dual of the field strength tensor is defined as $\tilde{F}_{\mu\nu} = \frac{1}{2} \epsilon_{\mu\nu\alpha\beta} F^{\alpha\beta}$. We do not consider the coupling of the pseudoscalar to two dark photons explicitly here, since it would only contribute to invisible $\phi$ decays, but will not lead to a sizeable displaced visible final-state signature in the small kinetic mixing limit. Via the interactions in Eqs.~\eqref{eq:lepton_couplings} and \eqref{eq:gauge_couplings}, the pseudoscalar $\phi$ can either decay visibly (to $\gamma\gamma$, $\gamma\gamma_D$ or $e^+e^-$) or invisibly (to $\nu\nu$), with the decay rates
\begin{gather}
\label{eq:boson_decay}
    \Gamma(\phi\to\gamma\gamma) 
    = \frac{g_{\phi\gamma\gamma}^2m_\phi^3}{64\pi}, \quad
    \Gamma(\phi\to\gamma\gamma_D) 
    = \frac{g_{\phi\gamma\gamma_D}^2m_\phi^3}{32\pi}
    \biggl(1-\frac{m_{\gamma_D}^2}{m_\phi^2}\biggr)^3, \\
\label{eq:fermion_decay}
    \Gamma(\phi\to\nu\nu) 
    = \sum_{i\leq j}^3 \frac{|c_\nu^{ij}|^2m_\phi}{8\pi(1 + \delta_{ij})} 
    = \frac{c_\nu^2 m_\phi}{16\pi}, \quad 
    \Gamma(\phi\to e^+e^-) 
    = \frac{c_e^2 m_\phi}{8\pi}
      \biggl(1-\frac{4m_e^2}{m_\phi^2}\biggr)^{1/2}.
\end{gather}
The effective interactions described above can be motivated from ultraviolet-complete models. In this context, neutrino mass generation mechanisms like the inverse seesaw~\cite{Mohapatra:1986bd} that employ symmetry-protected small lepton-number violation are known to induce unobservably low rates for standard $0\nu\beta\beta$ decay. This makes such scenarios particularly challenging to test and to establish the potential Majorana nature of neutrinos~\cite{Rodejohann:2011mu, Atre:2009rg, Bolton:2019pcu}. Such frameworks can naturally lead to an enhanced coupling of the Majoron to neutrinos compared with conventional high-scale seesaw mechanisms. On the other hand, sizeable effective couplings to di-photon, photon-dark photon, and charged leptons can arise via triangle loop diagrams~\cite{Quevillon:2019zrd} in the presence of heavy fields charged under electromagnetism and an Abelian dark gauge group, and via global lepton number assignments of the heavy neutral leptons participating in the neutrino mass generation. This can have clear implications for various realizations of axion-like particle scenarios~\cite{Jaeckel:2010ni} in connection with global lepton number symmetry. See the supplementary material for a discussion of relevant classes of ultraviolet-complete scenarios.

\section{Complementary Probes}
\label{sec:constraints}

Before discussing the resulting signal in double beta decay experiments, we here summarize existing constraints in other experimental and observational environments, focusing on masses $m_\phi\approx 1$~MeV, as relevant for our work.

\paragraph*{Kaon decay:} 
Neglecting the final-state masses, the three-body decay rate is given by $\Gamma(K^\pm \to e^\pm \nu_e \phi) = G_F^2 f_K^2 m_K^3 c_\nu^2/(768\pi^3)$. The coupling $c_\nu$ can be bounded by the uncertainty in the two-body decay~\cite{Blum:2014ewa, Pasquini:2015fjv}, $\Gamma(K^+\to e\nu_e)/\Gamma(K^+\to \mu\nu_\mu) = (2.488 \pm 0.010)\times 10^{-5}$~\cite{NA62:2012lny}, giving $c_\nu < 3.6 \times 10^{-3}$ at 90\%~CL. Other charged mesons, e.g. $\pi^+$ and $D^+$, and pseudoscalar production via $Z\to \nu_e \bar{\nu}_e \phi$ and $h\to \nu_e \bar{\nu}_e \phi$, provide less stringent bounds~\cite{Blum:2014ewa, Berryman:2018ogk, Altmannshofer:2026opc}.

\paragraph*{Electron anomalous magnetic moment $(g-2)_e$:} 
At the one-loop level, the pseudoscalar $\phi$ contributes to the negative shift, $\Delta a_e \approx -c_e^2 f(m_\phi^2 / m_e^2)/(16\pi^2)$, with $f(x) = 1 + 2x - x(x - 1)\ln x + 2x(x - 3)\Phi(x)$ and $\Phi(x) = \ln[(\sqrt{x} + \sqrt{x - 4})/2]/\sqrt{1 - 4/x}$ for $x > 4$ and $\arccos(\sqrt{x}/2)/\sqrt{4/x - 1}$ for $x < 4$. The deviation of the measured $(g-2)_e$ from Cs and Rb~\cite{Parker:2018vye, Morel:2020dww} and the SM prediction requires $-15.2 \times 10^{-13} < \Delta a_e < 6.7\times 10^{-13}$ and thus $c_e < 3.1\times 10^{-5}$ for $m_\phi = 1.5~\text{MeV}$.

\paragraph*{Neutrino scattering: Borexino, XENONnT, MiniBooNE:} 
Fits~\cite{Bansal:2022zpi} to solar neutrino scattering data for the process $\nu_e e^- \to \nu_e e^-\gamma_{(D)}$ at Borexino~\cite{Borexino:2017fbd} and XENONnT~\cite{XENON:2022ltv} yield the bounds $c_\nu g_{\phi\gamma\gamma'}/(1 + \delta_{\gamma\gamma'}) <  \{8.7, 3.5\}\times 10^{-10}~\text{MeV}^{-1}$ at 90\%~CL, respectively, for $m_\phi = 1$~MeV. The process $\nu_e A \to \nu_e A\gamma$ at MiniBooNE~\cite{MiniBooNE:2020pnu} also gives $c_\nu g_{\phi\gamma\gamma} < 4.0\times10^{-9}~\text{MeV}^{-1}$ at 90\%~CL~\cite{Bansal:2022zpi}. The process $\nu_e e^-\to \nu_e e^-$ at Borexino meanwhile constrains $c_\nu c_e < 3.0\times 10^{-11}$ at 90\%~CL for $m_\phi = 1.5$~MeV~\cite{Coloma:2022umy}, see also~\cite{Coloma:2022umy, DeRomeri:2024iaw, Blanco-Mas:2024ale}.

\paragraph*{Beam dump:} 
The coupling $g_{\phi\gamma\gamma}$ induces the Primakoff process $\gamma A \to \phi A$ in electron (E141~\cite{Riordan:1987aw,Dobrich:2017gcm}, E137~\cite{Bjorken:1988as} and NA64~\cite{NA64:2020qwq}) and proton (CHARM~\cite{CHARM:1985anb}, NuCal~\cite{Blumlein:1990ay, Blumlein:1991xh} and MiniBooNE~\cite{Capozzi:2023ffu}) beam dump experiments, followed by $\phi \to \gamma\gamma$ in the detector downstream. For $c_\nu = 0$, the most stringent bound from E137 is $g_{\phi\gamma\gamma} < 1.2 \times 10^{-8}~\text{MeV}^{-1}$ at 95\%~CL~\cite{Proceedings:2012ulb, Dobrich:2015jyk, Dolan:2017osp}. The impact of $\phi\to\nu\nu$ decays for $c_\nu \neq 0$ is to reduce $\text{Br}(\phi\to\gamma\gamma)$ and the $\phi$ decay length. As in~\cite{Bansal:2022zpi}, we therefore consider the combination of beam dump constraints as the condition $g_{\phi\gamma\gamma} > 1.2 \times 10^{-8}~\text{MeV}^{-1}/\text{Br}^{1/2}(\phi\to\gamma\gamma)$ as long as $\text{Br}(\phi\to\nu\nu) < 1/2$.

The coupling $g_{\phi\gamma\gamma_D}$ can instead lead to the production of $\phi$ via the decays of pseudoscalar and vector mesons produced in proton beam dumps, i.e., $P\to\gamma\gamma_D\phi$ and $V\to \gamma_D\phi$, followed by $\phi\to\gamma\gamma_D$~\cite{deNiverville:2018hrc, deNiverville:2019xsx, Gninenko:2026mgn}. In~\cite{Jodlowski:2023sbi}, this has been studied for $m_{\gamma_D}\ll m_\phi$, yielding the bound $g_{\phi\gamma\gamma_D} > 3.6 \times 10^{-7}~\text{MeV}^{-1}$ at 95\%~CL from CHARM. For $c_\nu \neq 0$, we consider $g_{\phi\gamma\gamma_D} > 3.6 \times 10^{-7}~\text{MeV}^{-1}/\text{Br}^{1/2}(\phi\to\gamma\gamma_D)$ for $\text{Br}(\phi\to\nu\nu) < 1/2$.

Lastly, the coupling $c_e$ induces electron bremsstrahlung $e^- A \to e^- A\phi$, followed by $\phi\to e^+e^-$ downstream, as searched for at KEK~\cite{Konaka:1986cb}, E141~\cite{Riordan:1987aw}, E137~\cite{Bjorken:1988as}, Orsay~\cite{Davier:1989wz}, E774~\cite{Bross:1989mp}, NA64~\cite{NA64:2021aiq} and MiniBooNE~\cite{Capozzi:2023ffu}. The strongest bound from E137 excludes an interval of $c_e$ values above $c_e > 3.8\times 10^{-8}$ at 95\% CL for $m_\phi = 1.5$~MeV. For $c_\nu \neq 0$, we consider the limit $c_e > 3.8\times 10^{-8}/\text{Br}^{1/2}(\phi\to e^+e^-)$ as long as $\text{Br}(\phi\to \nu\nu) < 1/2$. Another signature considered by the NA64 experiment~\cite{NA64:2021xzo} was $e^- A\to e^- A\phi$ followed by $\phi\to\text{inv.}$, setting the bound $c_e < 5.2\times 10^{-6}$ at 90\% CL for $m_\phi = 1.5$~MeV. For $c_\nu \neq 0$, we consider this bound as $c_e > 5.2\times 10^{-6}/\text{Br}^{1/2}(\phi\to\nu\nu)$, for $\text{Br}(\phi\to \nu\nu) > 1/2$.

\paragraph*{BaBar:} 
For $c_\nu \neq 0$, bounds on $g_{\phi\gamma\gamma}$, $g_{\phi\gamma\gamma_D}$ and $c_e$ can be found by recasting the BaBar dark photon search~\cite{BaBar:2017tiz}. For $g_{\phi\gamma\gamma}$ and $c_e$ ($g_{\phi\gamma\gamma_D}$), the mono-photon signature is provided by the process $e^+e^- \to \gamma \phi$ ($\gamma \gamma_D \phi$) followed by $\phi\to \gamma\gamma/e^+e^-$ ($\gamma\gamma_D$) outside the detector, or $\phi\to\nu\nu$ at any distance. To recast, we integrate the relevant cross section $d\sigma/dE_\gamma^*d\cos\theta_\gamma^*$ over the range $-0.4 < \cos\theta_\gamma^* < 0.6$ and $E_\gamma^* > 3$~GeV, and multiply by the probability of invisible $\phi$ decay $P_\text{inv} = 1 - P_{\text{vis}}$, with $P_{\text{vis}} = [1 - \text{Br}(\phi\to\nu\nu)][1 - \exp(-\Gamma_\phi L/\gamma\beta)]$, $L = 275$~cm, and the average boost factor $\langle\gamma\beta\rangle \approx 6400$. We require this to be less than the maximally allowed dark photon cross section ($e^+e^- \to \gamma\gamma_D$) with $m_{\gamma_D} = m_\phi$ and $\epsilon = 9.5\times 10^{-4}$, integrated over $-0.4 < \cos\theta_\gamma^* < 0.6$. The bounds, which depend on the value of $c_\nu$, reach down to $g_{\phi\gamma\gamma} < 1.6\times 10^{-7}~\text{MeV}^{-1}$ and $g_{\phi\gamma\gamma_D} < 2.6\times 10^{-6}~\text{MeV}^{-1}$ for $m_\phi = 1$~MeV and $c_e < 3.0\times 10^{-4}$ for $m_{\phi} = 1.5$~MeV, at 90\%~CL. For $g_{\phi\gamma\gamma}$, analogous but less stringent bounds can be obtained from the radiative decay $\Upsilon(nS) \to \gamma \phi$, constrained by $\Upsilon(nS) \to \gamma + \text{inv.}$~\cite{Wilczek:1977pj, Weinberg:1977ma, CrystalBall:1990xec, BaBar:2010eww, Dolan:2017osp}.

Another collider signal is $e^+e^- \to 3\gamma$, induced by $e^+e^-\to \gamma \phi$ followed by $\phi\to\gamma\gamma$ within the detector. We note that the two photons from $\phi$ decay can only be resolved for sufficiently large $m_\phi$; thus, bounds from Belle~II~\cite{Belle-II:2020jti} do not extend to $m_\phi = 1$~MeV. However, at LEP, the process $e^+e^- \to \gamma\gamma$ can set the bound $g_{\phi\gamma\gamma} < 1.4\times 10^{-4}~\text{MeV}^{-1}$ at 95\%~CL~\cite{Jaeckel:2015jla}. Because the $\phi\to\nu\nu$ decay decreases $\text{Br}(\phi\to\gamma\gamma)$ for $c_\nu \neq 0$, we exclude $g_{\phi\gamma\gamma} > 1.4\times 10^{-4}~\text{MeV}^{-1}/\text{Br}^{1/2}(\phi\to\gamma\gamma)$ for $\text{Br}(\phi\to\nu\nu) < 1/2$.

\paragraph*{Cosmology and Astrophysics:}
A pseudoscalar coupled to neutrinos and photons with $m_\phi = 1$~MeV can significantly alter the standard cosmological history, in particular the effective number of relativistic species $N_\text{eff}$ at the time of BBN and recombination. For $c_\nu \neq 0$ and $g_{\phi\gamma\gamma} = 0$, the requirement that $\nu\nu\to\phi$ does not keep $\phi$ in thermal equilibrium with the SM bath at the time of neutrino decoupling, $T \sim 1$~MeV, thereby contributing as $\Delta N_\text{eff} = 4/7$, enforces $c_\nu \lesssim 5\times 10^{-9}$~\cite{Huang:2017egl, Escudero:2019gvw}. At recombination, $\phi$ instead contributes to effective neutrino self-interactions $\nu\nu\to\nu\nu$, modifying the free streaming of neutrinos and impacting the CMB and large-scale structure formation~\cite{Hannestad:2004qu, Bell:2005dr, Oldengott:2017fhy}. Fits to cosmological data~\cite{Kreisch:2019yzn, Barenboim:2019tux, Camarena:2024daj} constrain $c_\nu \lesssim 10^{-2}$ and intriguingly also favour $c_\nu \sim \mathcal{O}(0.1)$, which relaxes the Hubble tension but is excluded by $0\nu\beta\beta\phi$ and meson decays~\cite{Blinov:2019gcj, Deppisch:2020sqh}. For $g_{\phi\gamma\gamma} \neq 0$ and $c_\nu = 0$, $\phi$ can thermalize via Primakoff scattering $\gamma q\to \phi q$ and modify $N_\text{eff}$ and the baryon-to-photon ratio $\eta$ via the decay $\phi\to \gamma\gamma$ after neutrino decoupling. Combining the BBN and CMB measurements of $N_\text{eff}$ excludes all relevant $g_{\phi\gamma\gamma}$ values for $m_\phi = 1$~MeV~\cite{Cadamuro:2011fd, Millea:2015qra, Depta:2020wmr}. We note, however, that the results above assume a standard cosmological evolution in the presence of $\phi$. For example, the presence of additional degrees of freedom, a non-zero neutrino asymmetry or a low reheating temperature can relax the bounds on $g_{\phi\gamma\gamma}$~\cite{Depta:2020wmr}. The interplay of effects when both $c_\nu$ and $g_{\phi\gamma\gamma}$ are non-zero may also be non-trivial and important, requiring a dedicated study. Similar constraints from BBN and CMB apply for the electron coupling $c_e$~\cite{Ghosh:2020vti}. No analysis has yet been performed for the impact of the dark photon coupling $g_{\phi\gamma\gamma_D}$.

The pseudoscalar couplings considered in this work are also subject to bounds from astrophysical phenomena, namely supernovae (SN) and binary neutron star (NS) mergers. For example, for $c_\nu, g_{\phi\gamma\gamma}\neq 0$, the Primakoff scattering $\gamma\to\phi$ and coalescence processes $\gamma\gamma\to\phi$ and $\nu\nu\to\phi$ produce $\phi$ in SN, enhancing the cooling rate if $\phi$ decays beyond the neutrino sphere. Requiring that the cooling does not exceed the measured neutrino luminosity for SN1987A excludes $9\times 10^{-10} < c_\nu < 2\times 10^{-5}$ for $g_{\phi\gamma\gamma} = 0$ and $10^{-12} < g_{\phi\gamma\gamma}/\text{MeV}^{-1} < 2\times 10^{-8}$ for $c_{\nu} = 0$~\cite{Bansal:2022zpi}, see also~\cite{Lucente:2020whw, Caputo:2022rca, Caputo:2022mah, Fiorillo:2022cdq}. Bounds on $g_{\phi\gamma\gamma}$ from the NS merger GW170817 are comparable~\cite{Diamond:2023cto, Dev:2023hax}. Possible $\gamma$-ray signatures from SN1987A~\cite{Hoof:2022xbe, Muller:2023vjm} and the diffuse SN background~\cite{Calore:2020tjw} via $\phi\to\gamma\gamma$ decay provide even more stringent bounds, $g_{\phi\gamma\gamma} < 10^{-14}~\text{MeV}^{-1}$ and $5\times 10^{-13}~\text{MeV}^{-1}$, respectively, but are relaxed for $c_\nu \neq 0$ and the presence of $\phi\to\nu\nu$ decays.

\section{Displaced energy deposition in double beta decay}
\label{sec:dbd}

For $m_\phi < Q$, where $Q$ is the isotope-dependent energy release ($Q$-value), we consider the emission of an on-shell pseudoscalar $\phi$. The differential decay rate for $0\nu\beta\beta\phi$ in the total electron kinetic energy $T$ is
\begin{align}
\label{eq:majoron_emission}
    \frac{d\Gamma_\phi}{dT} = 
    \frac{\kappa_{0\nu}}{8\pi^2}
    \left(\frac{m_e}{2R}\right)^2
    \left|\mathcal{M}_{0\nu}\right|^2 
    c_\nu^2 g(T) \sqrt{(Q - T)^2 - m_\phi^2},
\end{align}
where $\mathcal{M}_{0\nu}$ and $R$ are the nuclear matrix element (NME) and radius of the relevant isotope, respectively and we define, for convenience, $\kappa_{0\nu}\equiv G_F^4\cos^4\theta_C/8\pi^5 m_e^2$, where $G_F$ is the Fermi constant and $\theta_C$ is the Cabibbo angle. The function $g(T)$ arises from integrating over the phase space of $\phi$ and the energy difference of the electrons and is calculated in App.~B.

The observed signal in a double beta decay experiment depends on an interplay of the lifetime of $\phi$, its branching fractions to invisible and visible final states, and the dimensions and properties of the detector. If $\phi$ decays completely invisibly or has a proper decay length much longer than the detector size, the total \textit{visible} energy deposited in the detector is only provided by the $\beta\beta$ electrons. This corresponds to massive Majoron emission~\cite{Hirsch:1995in, Blum:2018ljv, Boudjema:2025okq, PandaX:2025tls, deVries:2025hqa}. From the non-observation of distortions to the SM $2\nu\beta\beta$ spectrum at KamLAND-Zen ($^{136}$Xe) yields upper limits on the coupling $c_\nu < 10^{-4}$ and $2\times 10^{-4}$ at 90\%~CL for $m_\phi = 1$~MeV and 1.5~MeV, respectively~\cite{Boudjema:2025okq}.

\begin{figure}[t!]
    \centering
    \includegraphics[width=0.70\textwidth]{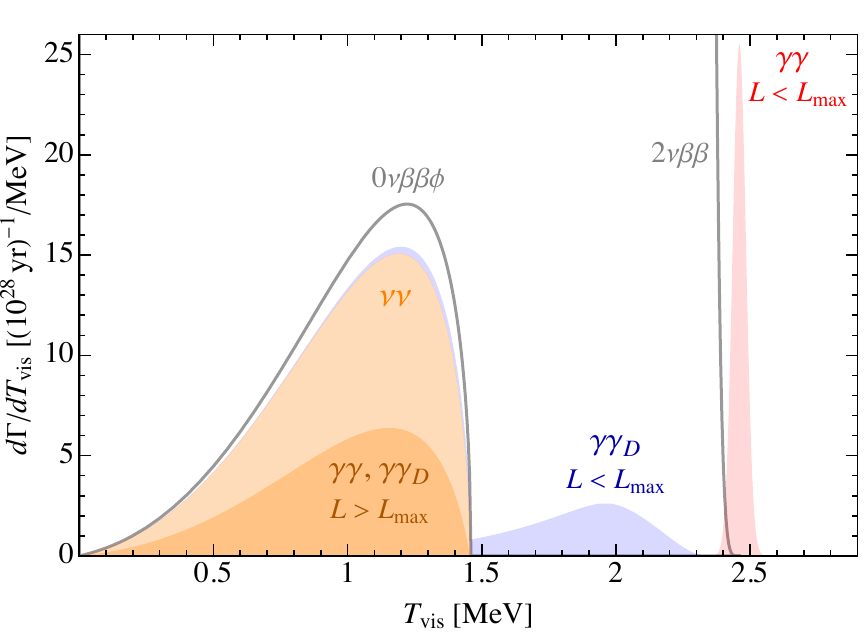}
    \caption{Differential double beta decay rate with respect to the total visible energy $T_\text{vis}$ for $^{136}$Xe and a pseudoscalar mass $m_\phi = 1$~MeV. Shown are stacked contributions from $\phi\to\nu\nu$ at any distance (light orange), $\phi\to\gamma\gamma_{(D)}$ beyond $L_\text{max} = 0.5$~m (dark orange), and fully visible (red) or partially visible (blue) decays within $L_\text{max}$. To visualize the energy release at the endpoint, we use an arbitrary Gaussian width of $25$~keV. SM $2\nu\beta\beta$ decay and massive Majoron emission $0\nu\beta\beta\phi$ and are shown for comparison in grey. The relevant couplings are chosen to be $c_\nu = 1.58 \times 10^{-6}$, $g_{\phi\gamma\gamma} = \sqrt{2}g_{\phi\gamma\gamma_D} = 3.15 \times 10^{-6}$~MeV$^{-1}$ so that $\text{Br}(\phi \to \nu\nu) = \text{Br}(\phi \to \gamma\gamma_{(D)}) = 1/2$ with the corresponding proper decay length $L_\phi = 2$~m.}
\label{fig:Tvisspectrum}
\end{figure}

However, for non-zero values of $g_{\phi\gamma\gamma}$, $g_{\phi\gamma\gamma_D}$ or $c_e$, the pseudoscalar can decay inside the detector, adding to the visible energy. As depicted for $\phi\to\gamma\gamma$ in Fig.~\ref{fig:diagram}, this may occur at a macroscopic distance $L$ from the $0\nu\beta\beta\phi$ decay process, resulting in a \textit{displaced and delayed} deposit of energy. The distribution in the total electron kinetic energy for the decay $\phi\to f$ ($f = \nu\nu, \gamma\gamma, \gamma\gamma_D, e^+e^-$) in a distance range $L_1 < L < L_2$ is
\begin{align}
\label{eq:majoron_emission_decay}
    \left.\frac{d\Gamma_f}{dT}\right|_{L_1}^{L_2} 
    = \frac{d\Gamma_\phi}{dT}
    \left[
          \exp\left(-\frac{L_1}{\gamma\beta L_\phi}\right) 
        - \exp\left(-\frac{L_2}{\gamma\beta L_\phi}\right)
    \right] \times \text{Br}(\phi\to f),
\end{align}
where $L_\phi = 1/\Gamma_\phi$ is the proper $\phi$ decay length and the boost factor is $\gamma\beta = [(Q - T)^2/m_\phi^2 - 1]^{1/2}$. Note that the presence of the decay probability in the square brackets, depending on the $\phi$ energy $E_\phi = Q - T$, will modify the shape of the spectrum from the canonical $0\nu\beta\beta\phi$ case in Eq.~\eqref{eq:majoron_emission}. 

Considering the possible decays of $\phi$ to $\nu\nu$ (always contributing but always invisible) and to one of $f = \gamma\gamma, \gamma\gamma_D, e^+e^-$ (visible if inside the detector), the distribution with respect to the total visible energy $T_\text{vis}$ is 
\begin{align}
\label{eq:majoron_emission_visible}
    \frac{d\Gamma_f}{dT_\text{vis}} 
    = \left.\frac{d\Gamma_{\nu\nu}}{dT}\right|_0^\infty 
    + \left.\frac{d\Gamma_f}{dT}\right|_{L_\text{max}}^\infty 
    + \left.\frac{d\Gamma_f}{dT_\text{vis}}\right|_0^{L_\text{max}},
\end{align}
with the maximal distance $L_\text{max}$ (size of the detector) within which the displaced energy release can be observed. Here, the first term corresponds to $\phi\to\nu\nu$ at any decay distance and the second term to $\phi\to f$ at $L > L_\text{max}$. These contribute to the \textit{bulk} of the visible energy spectrum with $T_{\text{vis}} = T$. In Fig.~\ref{fig:Tvisspectrum}, we show these as light and dark orange stacked spectra, respectively, for the $f = \gamma\gamma, \gamma\gamma_D$ scenarios. The $\nu\nu$ contribution has a shape identical to the $0\nu\beta\beta\phi$ spectrum, albeit suppressed by $\text{Br}(\phi\to\nu\nu)$. As noted, the $\gamma\gamma$ and $\gamma\gamma_D$ contributions are slightly distorted due to the $T$-dependent probability of decaying outside of $L_\text{max}$.

The third term in Eq.~\eqref{eq:majoron_emission_visible} represents the contribution of the visible decay inside the detector. The shape of this contribution depends on whether the decay mode is \textit{fully} or \textit{partially} visible. For the fully visible final states $f = \gamma\gamma, e^+e^-$, the total deposited energy adds up to the $Q$-value, and the theoretical spectrum is simply a Dirac delta function $\propto \delta(T_\text{vis} - Q)$. In Fig.~\ref{fig:Tvisspectrum}, we visualize this as the red Gaussian peak, representing a finite detector resolution. In the $e^+e^-$ case, the positron typically releases all of its kinetic energy through inelastic collisions with atomic electrons (ionization and excitation) before annihilating with an electron; see Fig.~\ref{fig:diagram}. For the partially visible final state $f = \gamma\gamma_D$, only $\gamma$ adds to the visible energy, $T_\text{vis} = T + E_\gamma$. This results in a contribution to the bulk spectrum, albeit peaking at a higher energy than $0\nu\beta\beta\phi$ and closer to the endpoint; see the blue shaded contribution $\gamma\gamma_D$ in Fig.~\ref{fig:Tvisspectrum}. 

Note that, generally speaking, the visible decays of $\phi$ can be either prompt, i.e., so close to the primary double beta decay that they are experimentally not resolvable, or displaced. In the latter case, specific types of detectors will be able to determine the presence of a displaced vertex, either through their spatial or temporal resolution. This may also allow measuring the prompt, $T_\text{prompt} = T$, and displaced, $T_\text{displ} = T_\text{vis} - T$ energy releases. In the example shown in Fig.~\ref{fig:Tvisspectrum}, the couplings are so small that the proper decay length of $\phi$ is $L_\phi = 2$~m, and the visible decays will have a large displacement.

Correlations among the kinematic variables from the on-shell $\phi$ production followed by a boosted two-body decay are expected to allow pinpointing the specific mechanism and reduce background. For example, for fully visible $\phi$ decay modes, the prompt and displaced energy releases are simply related as $Q = T_\text{prompt} + T_\text{displ}$ and the particles produced at the displaced vertex typically point away from the primary double beta vertex. More details on the distributions above and for additional kinematic variables such as the total $\phi$ energy $E_\phi$, single photon energy $E_\gamma$ and the angle between two outgoing photons can be found in App.~B.

\section{Discussion and conclusions}
\label{sec:sensitivity}

In order to asses the sensitivity of double beta decay experiments, we consider a simplified setup. We assume that the primary double beta decay occurs at the center of a spherical detector of radius $L_\text{max}$ within which any energy release is registered without loss. If a decay occurs beyond a distance $L_\text{min}$, we assume that the detector can identify it as a displaced vertex. We consider such a signal to be background-free. In addition, there are two other signal contributions as discussed above: Prompt ($L < L_\text{min}$) and fully visible decays $\phi\to\gamma\gamma$, $\phi\to e^+e^-$, which will mimic standard $0\nu\beta\beta$ decay with energy release at the $Q$-value. Here, we consider a typical background level for $0\nu\beta\beta$ searches. Finally, invisible decays $\phi\to\nu\nu$ and decays outside the detector contribute to a continuous spectrum. Here, we use the statistical approach in \cite{Boudjema:2025okq} to model the deviation of the massive Majoron-like signature from the SM $2\nu\beta\beta$ spectrum, including uncertainties in the NMEs. In order to understand the understand the sensitivity of both existing and future searches, we use three experimental configurations, all using $^{136}$Xe as double beta decay isotope: (a) a configuration based on KamLAND-Zen \cite{KamLAND-Zen:2024eml}, representing existing searches, with a nominal $0\nu\beta\beta$ sensitivity of $T^{0\nu}_{1/2} = 10^{26}$~yrs and $L_\text{max} = 4$~m; (b) a configuration based on nEXO \cite{nEXO:2021ujk}, representing current efforts, with $T^{0\nu}_{1/2} = 10^{28}$~yrs and $L_\text{max} \approx 19$~m; (c) and a far-future configuration where we scale nEXO by two orders of magnitude to $T^{0\nu}_{1/2} = 10^{30}$~yrs and $L_\text{max} \approx 86$~m. For simplicity, we use a universal minimal distance $L_\text{min} = 0.1$~m beyond which displaced vertices can be detected. The details of the experimental configurations and the statistical procedure employed are presented in App.~C.

While the above assumptions are very simplified and a realistic sensitivity analysis will require detector-specific simulations of the geometry, material properties and background, they are intended to cover typical double beta decay experiments. For example, segmented bolometer arrays such as CUORE~\cite{CUORE:2024ikf, CUORE:2024fak, Zhao:2025iyx} and CUPID \cite{CUPID-0:2022yws, CUPIDInterestGroup:2019inu, Zhao:2025iyx} are naturally expected to be sensitive to displaced energy release. Likewise, liquid-scintillator detectors such as KamLAND-Zen \cite{KamLAND-Zen:2024eml}, SNO+ \cite{SNO:2021xpa, Inacio:2024yfm} and JUNO \cite{Zhao:2016brs, Liu:2018fpq}. as well as time projection chamber (TPC) detectors like PandaX-4T~\cite{PandaX:2024fed, Qian:2025kxb} and PandaX-III \cite{Zhang:2023ywy} provide large detection volumes potentially suitable to detect displaced vertices. With existing searches ($T^{0\nu}_{1/2} \approx 10^{30}$~yrs) starting to probe inversely ordered light Majorana neutrinos and currently planned experiments ($T^{0\nu}_{1/2} \approx 10^{28}$~yrs) designed to cover the inverse range completely, there are already efforts to plan for more sensitive experiments to probe normally ordered neutrinos with $T^{0\nu}_{1/2} \approx 10^{30}$~yrs. 

\begin{figure*}[t!]
    \centering
    \includegraphics[width=0.49\textwidth]{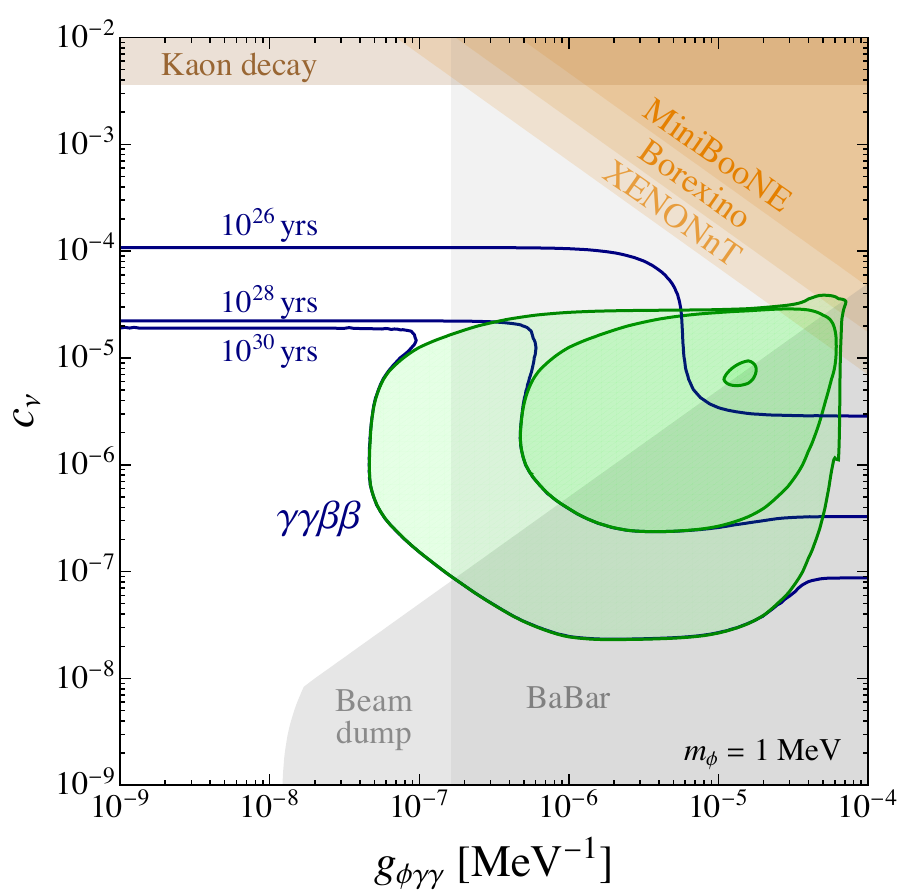}
    \includegraphics[width=0.49\textwidth]{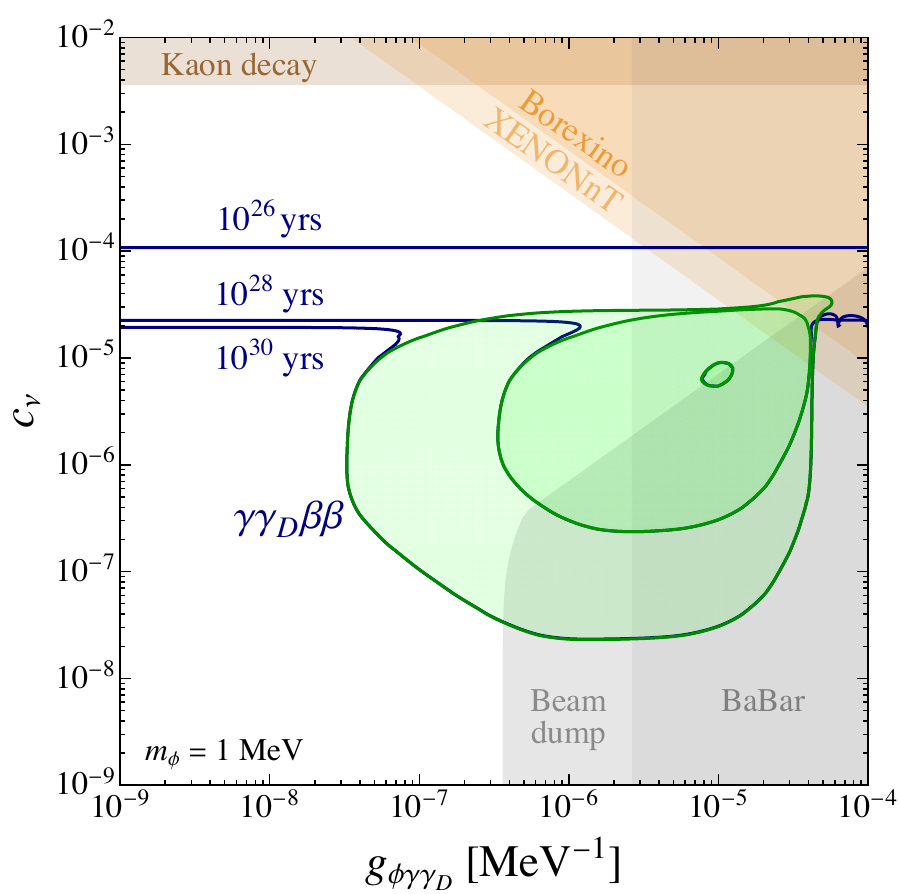}\\
    \includegraphics[width=0.49\textwidth]{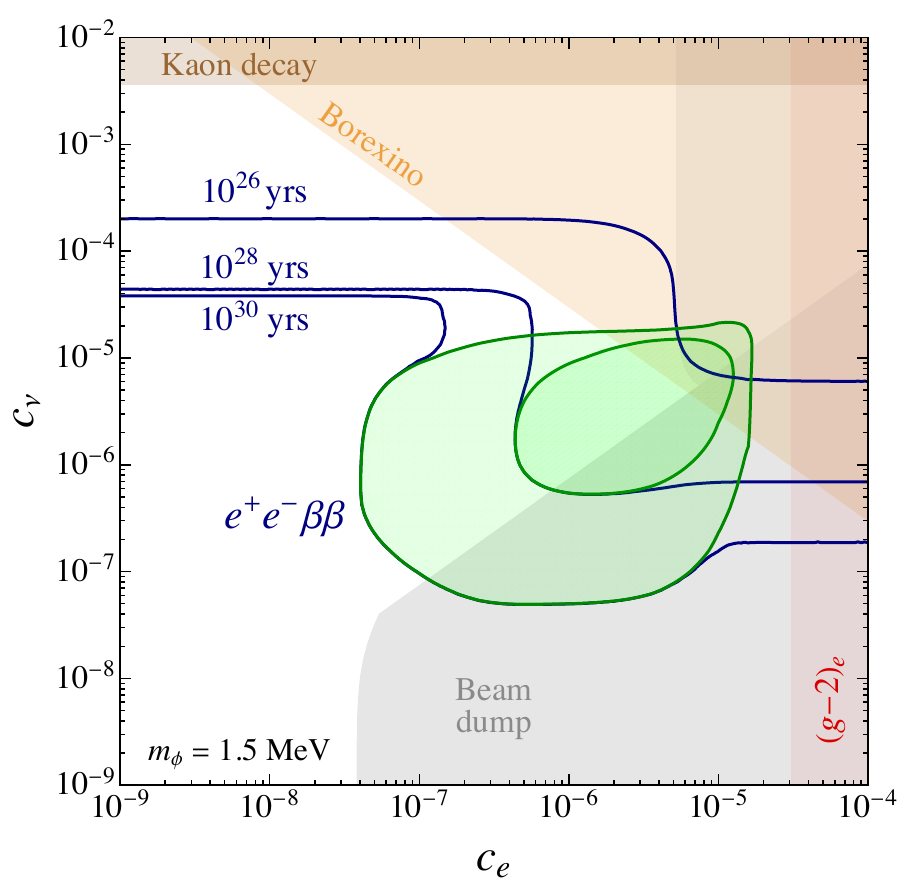}
    \caption{Sensitive parameter regions for overall exclusion (blue curves) and displaced signal exclusion (green areas) at 90\%~CL for $\gamma\gamma\beta\beta$ ($m_\phi = 1$~MeV, top-left), $\gamma\gamma_D\beta\beta$ ($m_\phi = 1$~MeV, top-right) and $e^+e^-\beta\beta$ ($m_\phi = 1.5$~MeV, bottom). The three contours and regions correspond to different, simplified experimental setups using $^{136}$Xe: KamLAND-Zen, with the current $0\nu\beta\beta$ exclusion $T_{1/2}^{0\nu} > 10^{26}$~yr, nEXO, with projected sensitivity $T_{1/2}^{0\nu} > 10^{28}$~yr and n$^2$EXO, a hypothetical far-future setup with projected sensitivity $T_{1/2}^{0\nu} > 10^{30}$~yr. Details of the experimental setups as well as the complementary constraints, indicated by the shaded regions, are described in the text.}
\label{fig:sensitivity}
\end{figure*}
Fig.~\ref{fig:sensitivity} illustrates the sensitive parameter regions at 90\%~CL for the three displaced modes considered, $\gamma\gamma\beta\beta$ ($m_\phi = 1$~MeV, top-left), $\gamma\gamma_D\beta\beta$ ($m_\phi = 1$~MeV, top-right) and $e^+e^-\beta\beta$ ($m_\phi = 1.5$~MeV, bottom). In each case, we turn on one of the couplings to visible particles, $g_{\phi\gamma\gamma}$, $g_{\phi\gamma\gamma_D}$, $c_e$, with the other two set to zero. The plots present the overall exclusion limits, incorporating all three signatures (invisible decays or outside the detector, background-free displaced signal, prompt energy release at $Q$-value), delineated by blue curves. Alongside, the 90\% CL exclusion based on the background-free displaced signal is indicated by the green shaded areas. As introduced above, the exclusions are determined by assuming non-observation of any of the signals at the three simplified experimental configurations representing current ($10^{26}$~yrs), near-future ($10^{28}$~yr) and far-future ($10^{30}$~yr) sensitivities. Constraints from other terrestrial experiments are overlaid as indicated. As can be seen, these mainly exclude large values of couplings to visible particles but the parameter space with $\text{Br}(\phi\to\nu\nu) \gtrsim 1/2$ is testable in double beta decay experiments. The scenario $\phi\to\gamma\gamma$ will be difficult to probe, though, due to the existing bounds from BaBar and beam dump experiments, requiring far-future sensitivity. The sensitivity to $c_\nu$ for small $g_{\phi\gamma\gamma}$, $g_{\phi\gamma\gamma_D}$, $c_e$ arises from invisible decays or decays outside the detector, equivalent to massive Majoron emission. Note that there is little improvement from the $10^{28}$~yrs to $10^{30}$~yrs configurations. This is because the sensitivity is limited by the NME uncertainties rather than experimental statistics in this regime. On the other hand, the displaced signal remains highly sensitive if the couplings are sufficiently small, corresponding to a long-lived $\phi$. For large $g_{\phi\gamma\gamma}$ and $c_e$, $\phi$ will decay visibly but promptly, thereby mimicking the usual $0\nu\beta\beta$ energy release near the endpoint. This improves the sensitivity to $c_\nu$ as double beta experiments have a highly reduced background level here, but this regime is largely ruled out by other experiments.

In this paper, we propose novel displaced modes of neutrinoless double beta decay, featuring characteristic displaced energy deposition signals that can be searched for in current and upcoming experiments. The long-lived particle in question is a light Majoron-like pseudoscalar, produced on-shell in $0\nu\beta\beta\phi$ decay. We then consider its subsequent decays as $\phi\gamma\gamma$, $\phi\to\gamma\gamma_D$ and $\phi\to e^+e^-$. The necessary underlying effective couplings naturally arise in popular classes of symmetry-driven ultraviolet-complete models that cannot be probed by the conventional $0\nu\beta\beta$ decay searches. We demonstrate that double beta decay experiments are expected to be sensitive to such a long-lived pseudoscalar, complementing and improving upon existing constraints. This opens a novel pathway to probe light and weakly interacting particles through displaced signals in double beta decay experiments.


\begin{acknowledgments}
F.~F.~D. would like to thank Jianglai Liu, Shao-Feng Ge and Oleg Titov for useful discussions. F.~F.~D. would like to thank the Tsung-Dao Lee Institute, where part of this work was undertaken, for their hospitality. F.~F.~D. thanks the Institute for Nuclear Theory at the University of Washington for its hospitality and the US Department of Energy for partial support during the completion of this work. N.-I.~B. would like to thank the CUORE/CUPID group at the Yale Wright Laboratory, where part of this work was carried out, for useful discussions and generous hospitality. Fig.~\ref{fig:diagram} was created with the help of Gemini~3.1~Pro. P.~D.~B. is supported by the Slovenian Research Agency under the research core funding No. P1-0035 and in part by the research grants N1-0253 and J1-4389. F.~F.~D. and N.-I.~B. acknowledge support from the UK Science and Technology Facilities Council (STFC) via the Consolidated Grant ST/X000613. C.~H. is funded by the Generalitat Valenciana under Plan Gen-T via CDEIGENT grant No. CIDEIG/2022/16 and would also like to acknowledge support in part by the Spanish grants PID2023-147306NB-I00 and CEX2023-001292-S (MCIU/AEI/10.13039/501100011033).
\end{acknowledgments}


\appendix
\section{Ultraviolet-complete models} 
\label{sec:UV}

We discuss here how the effective interactions considered in the main text can be naturally relevant in classes of well-motivated symmetry-driven minimal frameworks and then present an interesting example of an ultraviolet-complete scenario.

\subsection{Enhanced couplings of Majoron to neutrinos, photons and dark photons}

If lepton number $L$ is associated with a global $U(1)_L$ symmetry, the spontaneous breaking of this symmetry can be associated with a pseudo-Nambu-Goldstone boson (PNGB), called the Majoron~\cite{Chikashige:1980ui, Schechter:1981cv}. Such a Majoron naturally couples to neutrinos, dynamically realizing the type-I seesaw via the vacuum expectation value of the global $U(1)_L$ symmetry breaking. In this minimal setup, known as the singlet Majoron Model, a SM singlet scalar
\begin{align}
\Phi = \frac{1}{\sqrt{2}}(f_N + \hat{s}) e^{iJ/f_N} \,,
\end{align}
with lepton number $L=-2$ couples to right-handed neutrinos (RHNs, SM gauge singlet fermions) $N_R$ as
\begin{equation}
\label{Lag:maj}
	\mathcal{L} \supset \frac{1}{2} \lambda_N \overline{N^C_R} \Phi N_R 
    + \text{h.c.},
\end{equation}
such that the RHN mass matrix is given by $m_N = \lambda_N f_N / \sqrt{2}$. The radial mode $\hat{s}$ is heavy and can be integrated out, and the Majoron $J$ is the PNGB. $J$ can obtain mass via a soft, global symmetry-breaking term, in the case of an anomalous $U(1)_L$ (in the presence of new heavy chiral fields), or a connection to gravity~\cite{Akhmedov:1992hi, Rothstein:1992rh}, making its mass a free parameter. Assuming three light neutrinos and $p$ heavy neutrinos diagonalized in the basis $(\nu^c, N_R) = V n_R$, such that $V$ is a $(3 + p)\times(3 + p)$ mixing matrix, the general tree-level coupling of the Majoron to the neutrinos can be expressed as~\cite{Pilaftsis:1993af}
\begin{equation}
\label{lag:maj-n}
	\mathcal{L} \supset 
    -i \frac{J}{2f_N} \sum_{i,j=1}^{3+p} \overline{n_i} 
    \left[
	   C_{ij} (m_i P_L -m_j P_R) + C_{ji} (m_j P_L -m_i P_R) 
      + \delta_{ij} \gamma_5 m_i
	\right] n_j,
\end{equation}
where $C_{ij} = \sum_{k=1}^3 V_{ki} V_{kj}^*$ and $m_i$ ($i = 1,2,\cdots,3+p$) correspond to the physical neutrino masses arranged in ascending order. In this standard seesaw setup, the compatibility with the observed light neutrino masses will require $f_N$ to be quite large for order-unity elements of $Y^\nu$, leading to couplings that are too suppressed to be of phenomenological interest\footnote{Without any additional ingredients, such a scenario generates Majoron couplings to quarks and charged leptons at the one-loop level. Couplings to photons and gluons are generated at the two-loop level and are expected to be further suppressed.}. 

Interestingly, in the presence of an enhanced symmetry that protects lepton number, a lower seesaw scale and correspondingly a lower value of $f_N$ become naturally relevant. An example is provided by the inverse seesaw \cite{Mohapatra:1986bd}, which also naturally motivates the search for the modes under discussion such as $\gamma\gamma\beta\beta$ and $\gamma\gamma_D\beta\beta$ decays, since the usual $0\nu\beta\beta$ decay can be unobservably rare \cite{Rodejohann:2011mu, Atre:2009rg, Bolton:2019pcu}. The most general relevant couplings in the inverse seesaw scenario are given by
\begin{equation}
\label{Lag:maj-2}
	\mathcal{L} \supset 
      \frac{1}{2} \kappa_{N_R}\overline{N^C_R} \Phi N_R 
    + \frac{1}{2} \kappa_{N_L}\overline{N^C_L} \Phi N_L + \text{h.c.},
\end{equation}
leading to a mass matrix of the form
\begin{equation}
\label{eq:ISS}
    \mathcal{M}_\nu = \begin{pmatrix}
        0 &m_D & 0 \\
        m_D^\top & \mu_R & m_N^\top \\
        0 & m_N & \mu_L
    \end{pmatrix},
\end{equation}
in the basis $(\nu^C, N, N^C)_R$, where $\mu_{L(R)} = \kappa_{L(R)} f_N/\sqrt{2}$. In this scenario, the smallness of the $\mu$ parameters $\mu_{L(R)}\ll m_D \ll m_N$ is ensured by restoration of lepton number symmetry in the vanishing $\mu$ limit and, consequently, $f_N$ can be naturally small, enhancing couplings to neutrinos $c^{ij}_\nu = -m_\nu^{ij}/f_N$.

A sizeable coupling of the Majoron with $\gamma\gamma$ can arise via a triangle loop diagram involving a heavy fermion multiplet in the loop which contains an electromagnetically charged (under $U(1)_Q$) component (enabling coupling to photons) and is charged under global $U(1)_L$ (enabling the coupling of the Majoron to the heavy fermion in the loop). Likewise, a coupling of the Majoron with $\gamma\gamma_D$ can arise if the heavy fermion multiplet above also carries a dark $U(1)_D$ charge (in addition to being charged under $U(1)_Q$ and global $U(1)_L$), or if there is a kinetic mixing term between photon-dark photon field strengths. Therefore, a minimal setup realizing both enhanced couplings to neutrinos and $\gamma\gamma$ ($\gamma\gamma_D$) arises from a heavy multiplet charged under global $U(1)_L$ and local $U(1)_Q$ $(U(1)_D)$.

To explore the coupling of $\phi$ with (dark) photons, it is instructive to consider a gauged $U(1)_D$ extension of the SM gauge group. Let us denote the associated gauge boson by $A_D^\mu$. We will remain agnostic regarding the mechanism generating their masses. In the broken phase of the SM, the generic renormalizable terms in the Lagrangian are
\begin{align}
    \mathcal{L} &\supset 
      \frac{1}{2}(\partial_\mu \phi)(\partial^\mu \phi) 
    - \frac{m_\phi^2}{2}\phi^2 
    - \frac{1}{4} F_{D\mu\nu} F_D^{\mu\nu} 
    + \frac{m_{\gamma_D}^2}{2} A_{D\mu} A_D^\mu  
    - \frac{\epsilon}{2} F_{\mu\nu} F_D^{\mu\nu} 
    - g_D j_\mu^D A_D^\mu,
\end{align}
where $g_D$ is the $U(1)_D$ gauge coupling and $\epsilon$ is the kinetic mixing between $A_D^\mu$ and the photon field $A^\mu$. Generically, $\epsilon \neq 0$ is generated in the presence of heavy degrees of freedom charged under $U(1)_D$ and $U(1)_Q$, e.g., for heavy fermions with masses $M_i$ and $U(1)_D$ charges $Q_i^D$,
\begin{align}
\label{eq:kinetic_mixing}
    \epsilon \approx 
    \frac{e g_D}{12\pi^2} \sum_i Q_i Q_i^D \ln \frac{\mu^2}{M_i^2}.
\end{align}
Here, $\mu$ is the renormalization scale. However, for masses $m_{\gamma_D} \lesssim 1~\text{GeV}$, stringent constraints on $\epsilon$ arise from various beam-dump and collider experiments. Thus, we take the kinetic mixing to be small in the following, $\epsilon \ll 1$. The current $j_\mu^D$ contains light degrees of freedom, which we assume can only lead to invisible decays of $\gamma_D$.

\subsection{Example ultraviolet-complete realisations}

\subsubsection{Model~I: Explicit inverse seesaw realisation}

An enhanced coupling of $\phi$ to neutrinos can be realized either at tree level or radiatively, via symmetry-protected small lepton-number violation associated with the generation of small neutrino masses. Interestingly, in the radiative realization, the heavy multiplet that can generate an effective coupling to a photon-(dark) photon pair can naturally serve as the anchor for neutrino mass generation. In what follows, we provide a minimal working example of such a realization in the context of the inverse seesaw. Of course, the scenario can be straightforwardly generalized to any well-motivated gauge extensions of the SM that implement the inverse seesaw, or adapted to other cases where the violation of the lepton number is protected by some symmetry. This motivates the physics case for the dedicated search for our proposed $\gamma\gamma\beta\beta$ and $\gamma\gamma_D\beta\beta$ decay modes, in addition to $0\nu\beta\beta$ decay, which is expected to be highly suppressed in such scenarios.

\begin{figure}[t!]
    \centering
    \includegraphics[width=0.5\linewidth]{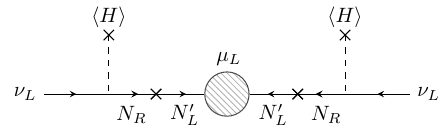}
    \caption{Realization of neutrino mass for the model examples, with the blob representing the loop generating the symmetry protected small $\mu_L$ parameter.}
    \label{fig:numass}
\end{figure}

\setlength{\tabcolsep}{10pt}
\renewcommand{\arraystretch}{1.2}
\begin{table}[t!]
\centering
\begin{tabular}{c|c|c|c|c|c}
\hline
  & $\Phi$ & $\Psi_{R(L)}$ & $S$ & $N_R$ & $N'_L$ \\
\hline
$SU(3)_C$ & 1 & $m$ & $m$ & 1 & 1 \\
$SU(2)_L$ & 1 & $n$ & $n$ & 1 & 1 \\
$U(1)_Y$  & 0 & 0   & 0   & 0 & 1 \\
\hline
$U(1)_{{B-L}}^\text{global}$ & $-1$ & $\frac{1}{2}$ ($-\tfrac{1}{2}$) & 
$-\tfrac{1}{2}$ & $-1$ & 1 \\
\hline
Spin & $0$ & $\tfrac{1}{2}$ & $0$ & $\tfrac{1}{2}$ & $\tfrac{1}{2}$ \\
\hline
\end{tabular}
\caption{Exotic field content of the Model~I, along with transformation properties under the SM gauge symmetries and global $U(1)_{B-L}$. Here, $m$ represents singlet or the adjoint representation of $SU(3)$: $m=1,8$ and $n$ is any non-trivial representation under $SU(2)$: $n=1,2,...$.}
\label{tab:m1}
\end{table}
Let us consider a model in which we extend the SM field content with the fields shown in Table~\ref{tab:m1}. The relevant new couplings for neutrino mass generation are
\begin{align}{\label{lag1:m1}}
    \mathcal{L} &\supset 
       y_\nu^{ik}\overline{L_{iL}} N_{kR} \tilde{H} 
     + m_N^{nk} \overline{(N'_{nL})^C}{N_{kR}} 
     + f_\Psi^{pq} \Phi \overline{(\Psi_{pR})^C}\Psi_{qR} \nonumber\\
    &+ h^{kp} S \overline{(N'_{kL})^C}(\psi_{pR})^C 
     +\mu_\Phi \Phi^\dagger S^2 + \text{h.c.},
\end{align}
where we assume the Einstein summation convention and the exact choice of the number of generations fixes the final sum upper limits for $k,n,p$. The index $i$ runs from 1 to 3, and we write the scalar as $\Phi = (f_\Phi +\hat{s}) e^{i\phi/f_N}/\sqrt{2}$. In this particular example under discussion, the neutrino mass matrix takes the form of Eq.~\eqref{eq:ISS} with $\mu_R = 0$ and $\mu_L$ is induced at the one-loop level with heavy $\Psi$ in the loop, cf. Fig.~\ref{fig:numass}. The general form for $\mu_L$ in this scenario is given by
\begin{equation}
\label{eq:ISSmu}
  {\bm{\mu_L}} \propto \sum_{k}\frac{ (\bm{h}\cdot m_{\Psi}^k \bm{I}_{k\times k} \cdot \bm{h^T})}{16\pi^2\sqrt{2}}  \left[\log\left(\frac{m_{\text{Re}(S)}^2}{{m_\Psi^k}^2}\right) 
  \frac{m_{\text{Re}(S)}^2}{m_{\text{Re}(S)}^2-{m_\Psi^k}^2} 
  - \left(m_{\text{Re}(S)}\to m_{\text{Im}(S)}\right)\right],
\end{equation}
where $\mu_L$ is in general a matrix fixed by the indices $k,n,p$ and the parentheses on the right-hand side contain a matrix product of the Yukawa coupling matrices which determines the corresponding entry. Without loss of generality, in the above we have used a basis in which $\Psi_R$ states are diagonal with the masses $m_{\Psi}^k= {f}_{\Psi}^{k} f_\Phi/\sqrt{2}$ and the splitting between real and imaginary components of $S$ is induced via $\mu_\phi$. We used proportionality to avoid specifying the exact numerical prefactors, which depend on the number of generations and the chosen color and $SU(2)$ representations of the new heavy fields. Note that the mass splitting between the real and imaginary components depends on the dimensionful coupling $\mu_\Phi$ and potential post-electroweak symmetry breaking corrections, which depend on the chosen $SU(2)$ representations. The generated light neutrino mass has the general form $m_\nu = m_D (m_N^{-1})^T \mu_L m_N^{-1} m_D^T$, leading to a coupling of the Majoron to neutrinos $c_\nu^{ij} = -i m_\nu^{ij} / f_\Phi$. As a benchmark, for $m_\nu \approx 0.1$~eV and $f_\phi \approx 1$~MeV, one can naturally obtain an enhanced value of interest, $c_\nu \approx 10^{-7}$, where the smallness of $\mu_L \propto f_\Phi$ motivates this choice. 

The couplings of $\phi$ to photons and gluons can arise via one-loop triangular graphs involving the heavy vector-like fermion $\Psi$, if it transforms non-trivially under $SU(2)_L$ and $SU(3)_C$, respectively. The relevant interactions are
\begin{equation}
\label{eq:phot_coup}
    \mathcal{L} \supset 
    - \frac{\alpha_2 \sin^2\theta_W}{4\pi} d_C^\Psi n_\psi T_L(R_\Psi) 
      \frac{\phi}{f_\phi} F_{\mu\nu} \tilde{F}^{\mu\nu} 
    - \frac{\alpha_s}{4\pi} d_L^\Psi n_\psi T_C(R_\Psi) 
      \frac{\phi}{f_\phi} G^a_{\mu\nu} \tilde{G}^{a,\mu\nu},
\end{equation}
where $\alpha_2$ and $\alpha_s$ are the weak and strong coupling constants, respectively. $d_C^\Psi$ and $d_L^\Psi$ are the dimensions for $SU(3)_C$ and $SU(2)_L$ transformations of $\Psi$, respectively, cf. Table~\ref{tab:m1}. $n_\psi$ is the number of $\Psi$ generations and $T_L(R_\Psi)$ and $T_C(R_\Psi)$ are the Dynkin indices of the $SU(2)_L$ and $SU(3)_C$ representations of $\Psi$, respectively. Above, we have also used the fact that two Weyl components of the vector-like $\Psi\equiv ((\Psi_L)_\alpha, (\Psi_R)^{\dagger\dot{\alpha}})^T$ are chirally charged under the $U(1)^{\text{global}}_{B-L}$. In the case where $\phi$ exhibits sizeable coupling to gluons, additional constraints become relevant~\cite{Blinov:2021say, Bauer:2021mvw, Chakraborty:2021wda}. For a numerical benchmark, taking the case of a single color-singlet fermion triplet case, i.e., $d_L^\Psi = 3$ and $d_C^\Psi = 1$, we obtain from Eq.~\eqref{eq:phot_coup} that $g_{\phi\gamma\gamma} = 2\alpha\sin^2\theta_W \pi^{-1}\approx 10^{-3}$ for $f_\phi = 1$~MeV, which is clearly large compared to the values we are interested in for a reasonably displaced signal. A more naturally viable and experimentally interesting case is obtained when $\phi$ couples to a photon-dark photon pair, as we describe in the following. Note that, similar to photons, a coupling to gluons can arise at the one-loop level via triangular graphs if $\Psi$ is charged non-trivially under $SU(3)_C$. In what follows, we will assume that $\Psi$ is a color-singlet $SU(2)$ state, thereby primarily playing the role of generating neutrino masses and enhancing the coupling of $\phi$ to neutrinos.

\begin{table}[t!]
\centering
\begin{tabular}{c||c|c|c|c|c|c|c}
\hline
& $\Phi$ & $\Phi'$ & $\Psi_{R(L)}$ & $\chi_{R(L)}$ & $S$ & $N_R$ & $N'_L$ \\
\hline
$SU(3)_C$ & 1 & 1 & 1 & 1 & 1 & 1 & 1 \\
$SU(2)_L$ & 1 & 1 & 1 & 1 & 1 & 1 & 1 \\
$U(1)_Y$  & 0 & 0 & 0 & $Y_\chi$ & 0 & 0 & 0 \\
\hline
$U(1)_D$  & 0 & 0 & 0 & $Q^D_\chi$ & 0 & 0 & 0 \\
$U(1)_{{B-L}}^\text{global}$ & $-1$ & 0 & $\tfrac{1}{2}$ ($-\tfrac{1}{2}$) & 
$l_\chi$ ($-l_\chi$) & $-\tfrac{1}{2}$ & -1 & 1 \\
$U(1)_X^\text{global}$ & 0 & -1 & 0 & $\tfrac{1}{2}$ ($-\tfrac{1}{2}$) & 
0 & 0 & 0\\
\hline
Spin & 0 & 0 & $\tfrac{1}{2}$ & $\tfrac{1}{2}$ & 0 & 
$\tfrac{1}{2}$ & $\tfrac{1}{2}$ \\
\hline
\end{tabular}
\caption{As in Table~\ref{tab:m1}, but showing the field content and quantum numbers in the extended Model~II. The gauge group $U(1)_D$ is necessary to realize the coupling of $\phi$ to $\gamma \gamma'$ and can be omitted if one is only interested in di-photon coupling.}
\label{tab:m2}
\end{table}

\subsubsection{Model~II: Extension for sizeable di-photon and photon-dark photon couplings}

To realize di-photon and photon-dark photon couplings in the ballpark of values relevant to the displaced signal and not too large (as in the previous minimal construction), we will consider the extended field content shown in Table~\ref{tab:m2}. We note that in this extension, a new $U(1)_X$ global symmetry is introduced to accommodate the di-photon and photon-dark photon final states. As in the previous discussion, $\Phi$ and $\Psi$ generate neutrino masses via the symmetry-protected inverse seesaw realization while also yielding sufficiently enhanced neutrino couplings. However, now that $\Psi$ is a complete SM singlet, it does not induce any coupling of $\phi$ to di-photons. Instead those couplings are generated by $\Phi' = (f'_\Phi +\hat{s})e^{i\phi'/f'_N}/\sqrt{2}$ and $\phi - \phi'$ mixing. Since we are also interested in the photon-dark photon final state, we will assume a gauged dark Abelian $U(1)_D$, which leads to a massive dark photon, as noted in the discussion preceding the model realization.

The heavy vector like fermionic multiplet $\chi$ charged under $U(1)_Y$ (and hence also under $U(1)_Q$) and $U(1)_D$ as $(Y_\chi, Q^D_\chi)$ leads to the relevant effective couplings of $\phi'$ to di-photon and photon-dark photon given by
\begin{align}
\label{eq:pdp_coup}
    \mathcal{L} \supset 
    &- \frac{\alpha_Y \cos^2\theta_W}{4\pi} n_\chi \frac{Y_\chi^2}{4} 
      \frac{\phi'}{f'_\phi} F_{\mu\nu} \tilde{F}^{\mu\nu} 
    - \frac{\sqrt{\alpha_Y \alpha_D} \cos\theta_W}{4\pi} n_\chi 
      \frac{Y_\chi Q^D_\chi}{2} \frac{\phi'}{f'_\phi} 
      F_{\mu\nu} \tilde{F}^{\mu\nu}_D \nonumber\\
    &- \epsilon \frac{\alpha_Y \cos^2\theta_W}{4\pi} n_\chi 
      \frac{Y_\chi^2}{4}\frac{\phi'}{f'_\phi} 
      F_{\mu\nu} \tilde{F}^{\mu\nu}_D.
\end{align}
Here, $\alpha_Y\equiv g_Y^2/4\pi$ and $\alpha_D\equiv g_D^2/4\pi$ are the coupling constants corresponding to hypercharge and the dark $U(1)_D$ gauge group. Assuming the kinetic mixing is small, we will neglect the third term in Eq.~\eqref{eq:pdp_coup}. For the benchmark choices $n_\chi = 1$, $Q^D_\chi = 1$, $Y = 2$, we then obtain the coupling of $\phi$ to photons given by $g_{\phi\gamma\gamma} = \theta_m \alpha_Y \cos^2\theta_W \pi^{-1}\approx 10^{-6}$ for $f_\phi' = 1$~GeV, where $\theta_m \leq 1/\sqrt{2}$ is the mixing factor between $\phi$ and $\phi'$. For the coupling of $\phi$ to photon-dark photon, we also obtain a similar value for $\alpha_D \sim \alpha_Y$. Therefore, we conclude that in this extended scenario all parts of the parameter space relevant to our displaced signal can be naturally realized for a reasonable choice of $f_\phi'$, while $f_\phi$ ensures sizeable couplings of $\phi$ to neutrinos. We note that if one is only interested in the di-photon couplings, the dark gauged $U(1)_D$ can simply be omitted.

We would like to stress that the resulting displaced $\gamma\gamma\beta\beta$ and $\gamma\gamma_D\beta\beta$ decays are clearly lepton number violating. However, this arises from the triangular effective interactions and is not tied to the neutrino mass generation, which is protected by the enhanced symmetry of the inverse seesaw, making the standard $0\nu\beta\beta$ decay extremely suppressed. While here we have discussed one example realization, the relevant idea can be straightforwardly generalized to other scenarios in which the violation of lepton number is protected by symmetry. One of the implications of this is that such scenarios (previously thought to be not probable by the standard $0\nu\beta\beta$ decay) can be probed using the proposed displaced mode in this work, provided that the Majoron presents the effective couplings discussed above, making it also very exciting for the axion-like particle scenarios~\cite{Jaeckel:2010ni}, with connection to lepton number symmetry.

\subsubsection{Sizeable dilepton couplings}

In the explicit model realizations above, we focused primarily on the scenario with enhanced di-photon and photon-dark photon couplings. In such a realization of the symmetry-protected inverse seesaw (cf. Table~\ref{tab:m1}), the coupling of $\phi$ to two electrons is proportional to the small symmetry-protected $\mu$ parameter with the relevant effective coupling $c_e \sim (\lambda_s/96\pi^2) m_e m_\nu/v^2$, which is highly suppressed~\cite{Herrero-Brocal:2023czw}. However, one can enhance such couplings by choosing a somewhat different realization of the inverse seesaw where $N_L$ carries zero lepton number, allowing for couplings of the form $\Phi \overline{N_R^C} N_L^C$ leading to an enhanced coupling to electrons~\cite{Vicente:2026zen}, $c_e \sim (y_\nu^2/16\pi^2) m_e /f_\phi$.
\section{Double beta decay distributions and rates}
\label{sec:distributions}

We derive the decay distributions used in this work, in particular, the differential rates for the $2\nu\beta\beta$, $\gamma\gamma\beta\beta$, $\gamma\gamma_D\beta\beta$ and $e^+e^-\beta\beta$ decay processes induced by the interactions of the pseudoscalar $\phi$ in the main text. We denote the four-momenta of the outgoing electrons as $p_{1,2} = (E_{1,2}, \vec{p}_{1,2})$ and the outgoing $\nu\nu$, $\gamma\gamma$, $\gamma\gamma_D$ or $e^+e^-$ as $p_{3,4} = (E_{3,4}, \vec{p}_{3,4})$.

In general, the double beta decay amplitudes can be factorized as
\begin{align}
    \mathcal{A}_f = \sum_n L_{\mu\nu}^f H^{\mu\nu}, \quad 
    f = \nu\nu, \gamma\gamma, \gamma\gamma_D, e^+e^-,
\end{align}
where $L_{\mu\nu}$ and $H^{\mu\nu}$ denote the leptonic and hadronic contributions, respectively, with the sum over intermediate nuclear states $n$. The hadronic part $H^{\mu\nu}$ is the same for each final state and equal to the hadronic part of the standard $0\nu\beta\beta$ decay process, due to the Majoron-like nature of the neutrino coupling to $\phi$. The leptonic parts are
\begin{align}
    L_{\mu\nu}^f &= 
    \frac{\mathcal{N}_{0\nu}}{p_\phi^2 - m_\phi^2 + i m_\phi \Gamma_\phi}
    \sum_{i,j} U_{ei} U_{ej} L_{\mu\nu}^{ij} V_f,
\end{align}
with
\begin{align}
    V_{\nu\nu} &= 
    \bar{u}_\nu(p_3)
    \left(c_\nu^{kl} P_L + c_\nu^{kl*}P_R\right)v_\nu(p_4), \nonumber\\
    V_{\gamma \gamma'} &= g_{\phi\gamma \gamma'}\epsilon_{\alpha\beta\rho\sigma} p_3^\alpha p_4^\beta \epsilon^{\rho*}(p_3)\epsilon^{\sigma*}(p_4), 
    \qquad \gamma' = \gamma, \gamma_D
    \nonumber\\
    V_{e^+e^-} &= -ic_e\bar{u}_e(p_3)
    \gamma_5 v_e(p_4).
\end{align}
Here, $\mathcal{N}_{0\nu} = G_F^2\cos^2\theta_C/\sqrt{4\pi R^2}$ is an appropriate normalisation factor with the nuclear radius $R = 1.3~\text{fm}\times A^{1/3}$, and $U$ is the PMNS mixing matrix. The momentum of the $s$-channel pseudoscalar is $p_\phi = p_3 + p_4$. The remaining part of the leptonic matrix element, containing the neutrino propagators and neutrino coupling to $\phi$, is given by
\begin{align}
    L_{\mu\nu}^{ij} &= \bar{u}_e(p_1)\gamma_\mu P_L 
    \frac{\slashed{q} + m_i}{q^2 - m_i^2}
    \left(c_\nu^{ij} P_L + c_\nu^{ij*}P_R\right)
    \frac{\slashed{q} - \slashed{p}_\phi + m_j}{(q - p_\phi)^2 - m_j^2}
    \gamma_\nu P_R v_e(p_2) \nonumber\\
    &= \frac{g_{\mu\nu}}{q^2} c_\nu^{ij*} \bar{u}_e(p_1) P_R v_e(p_2),
\end{align}
where in the second step we use that $H^{\mu\nu}$ is symmetric under $\mu\leftrightarrow\nu$ and that the momentum exchange through the neutrino propagators is large, $|q^2|\approx (100~\text{MeV})^2 \gg |p_\phi^2|, m_i^2$.

We thus obtain the differential rate
\begin{gather}
    d\Gamma_f = \left|\mathcal{M}_{0\nu}\right|^2 d\mathcal{G}_f,
\end{gather}
where $\mathcal{M}_{0\nu}$ is the nuclear matrix element (NME) for the standard $0\nu\beta\beta$ decay process and $d\mathcal{G}_f$ contains the leptonic contributions and differential phase space for the final state $f$;
\begin{align}
    d\mathcal{G}_{\nu\nu} &= 
    \sum_{k\leq l} 
    \frac{8\mathcal{N}_{0\nu}^2}{(p_\phi^2 - m_\phi^2)^2 + m_\phi^2 \Gamma_\phi^2}
    \frac{\big|\sum_{i,j}U_{ei}U_{ej}c_\nu^{ij*} c_\nu^{kl}\big|^2}{1+\delta_{kl}} (p_1 \cdot p_2)(p_3\cdot p_4) F^2(E_1, E_2) d\Pi_4 \nonumber\\
    &= \frac{4\mathcal{N}_{0\nu}^2}{(p_\phi^2 - m_\phi^2)^2 + m_\phi^2 \Gamma_\phi^2}
    c_\nu^4 (p_1 \cdot p_2)(p_3\cdot p_4) 
    F^2(E_1, E_2) d\Pi_4,
\label{eq:phase_space_factors-nunu}
\end{align}
for $f = \nu\nu$,
\begin{align}
    d\mathcal{G}_{\gamma\gamma'} &= 
    \frac{4\mathcal{N}_{0\nu}^2}{(p_\phi^2 - m_\phi^2)^2 + m_\phi^2 \Gamma_\phi^2} 
    \frac{\big|\sum_{i,j} U_{ei} U_{ej} c_\nu^{ij*} g_{\phi\gamma\gamma'}\big|^2}{1+\delta_{\gamma\gamma'}} (p_1 \cdot p_2)
    (p_3\cdot p_4)^2 
    F^2(E_1, E_2) d\Pi_4 \nonumber\\
    &= \frac{4\mathcal{N}_{0\nu}^2}{(p_\phi^2 - m_\phi^2)^2 + m_\phi^2 \Gamma_\phi^2}
    \frac{c_\nu^2 g_{\phi\gamma\gamma'}^2}{1+\delta_{\gamma\gamma'}}
    (p_1 \cdot p_2)(p_3\cdot p_4)^2 
    F^2(E_1, E_2) d\Pi_4,
\label{eq:phase_space_factors-gg}
\end{align}
for $f = \gamma\gamma'$, and
\begin{align}
    d\mathcal{G}_{e^+e^-} &= 
    \frac{8\mathcal{N}_{0\nu}^2}{(p_\phi^2 - m_\phi^2)^2 + m_\phi^2 \Gamma_\phi^2}
    \bigg|\sum_{i,j}U_{ei}U_{ej}c_\nu^{ij*} c_e\bigg|^2 (p_1 \cdot p_2)\Big[(p_3\cdot p_4) + m_e^2\Big] F^4(E_1, E_2, E_3, E_4) d\Pi_4 \nonumber\\
    &= 
    \frac{8\mathcal{N}_{0\nu}^2}{(p_\phi^2 - m_\phi^2)^2 + m_\phi^2 \Gamma_\phi^2}
    c_\nu^2 c_e^2 (p_1 \cdot p_2)\Big[(p_3\cdot p_4) + m_e^2\Big]
    F^4(E_1, E_2, E_3, E_4) d\Pi_4,
\label{eq:phase_space_factors-ee}
\end{align}
for $f = e^+e^-$. Eqs.~\eqref{eq:phase_space_factors-nunu} and~\eqref{eq:phase_space_factors-gg} include a symmetry factor of $1/2$ for identical final state particles. The four-body phase space element is
\begin{align}
\label{eq:dlips}
    d\Pi_4 = \left[\prod_{i = 1}^4 \frac{d^3\vec{p}_i}{(2\pi)^3 2 E_i}\right]
    \delta(Q - T - E_3 - E_4).
\end{align}
Working in the $s$-wave approximation for the outgoing electron pair, we include in Eqs.~\eqref{eq:phase_space_factors-nunu} and \eqref{eq:phase_space_factors-gg} the factor $F^2(E_1, E_2) \equiv F_0(Z_f, E_1)F_0(Z_f, E_2)$, with the Fermi function
\begin{align}
    F_0(Z_f, E_i)=  
    4\frac{(2 |\vec{p}_i| R)^{2(\gamma_0 -1)}}{\Gamma^2(1+2 \gamma_0)} e^{\pi y} 
    \left|\Gamma(\gamma_0 + i y)\right|^2,
\end{align}
where $\Gamma(x)$ is the Gamma function, $\gamma_0 = \sqrt{1 - (Z_f\alpha)^2}$, and $y = \alpha Z_f E_i/|\vec{p}_i|$, with the fine-structure constant $\alpha = 1/137$, the charge number of the final nucleus $Z_f$ and $|\vec{p}_i| = \sqrt{E_i^2 - m_e^2}$. In Eq.~\eqref{eq:phase_space_factors-ee}, if $\phi$ decays within the nucleus, the factor $F^4(E_1, E_2, E_3, E_4)$ should include the Fermi functions also for the emitted $e^+e^-$ pair, i.e.,
\begin{align}
    F^4(E_1, E_2, E_3, E_4) 
    = F_0(Z_f, E_1)F_0(Z_f, E_2)F_0(-Z_f, E_3)F_0(Z_f, E_4).
\end{align}
However, in the scenario where $\phi$ decays away from the nucleus, there is a negligible impact, with the plane-wave approximation suitable for the $e^+e^-$, i.e., $F^4(E_1, E_2, E_3, E_4) \approx F^2(E_1, E_2)$. In Eqs.~\eqref{eq:phase_space_factors-nunu}-\eqref{eq:phase_space_factors-ee}, we consider electron flavor couplings, $c_\nu^{ij} = i U_{ei}U_{ej}c_\nu$ and $c_e^{ij} = i\delta_{ei}\delta_{ej}c_e$, and use the unitarity of the PMNS mixing matrix to simplify the sum over neutrino mass eigenstates.

We first perform the integration over the phase space of the final state neutrinos, photons, or an electron-positron pair if $\phi$ is sufficiently displaced, writing
\begin{align}
    I_{\nu\nu}(T) &= 
    \int \left[\prod_{i = 3}^4 \frac{d^3\vec{p}_i}{(2\pi)^3 2 E_i}\right] 
    \frac{(p_3 \cdot p_4)}{(p_\phi^2 - m_\phi^2)^2 + m_\phi^2 \Gamma_\phi^2} 
    \delta(Q - T - E_3 - E_4) \nonumber\\
    &= \frac{1}{4(2\pi)^4} \int_0^{Q - T} dE_3 
    \left\{\frac{m_\phi}{2\Gamma_\phi}
    \arctan\left(\frac{\zeta_+ - \zeta_-}{1 + \zeta_+\zeta_-}\right) 
    + \frac{1}{4}\log\left(\frac{\zeta_+^2 + 1}{\zeta_-^2 + 1}\right)\right\},
\label{eq:integrals-nunu}
\end{align}
for $f = \nu\nu$,
\begin{align}
    I_{\gamma\gamma'}(T) &= 
    \int \left[\prod_{i = 3}^4 \frac{d^3\vec{p}_i}{(2\pi)^3 2 E_i}\right] 
    \frac{(p_3\cdot p_4)^2}{(p_\phi^2 - m_\phi^2)^2 + m_\phi^2 \Gamma_\phi^2}
    \delta(Q - T - E_3 - E_4) \nonumber\\
    &= \frac{1}{4(2\pi)^4} \int_{0}^{Q - T - m_{\gamma'}} dE_3 
    \left\{|\vec{p}_3||\vec{p}_4| + \frac{(m_\phi^2 -  m_{\gamma'}^2)^2 - m_\phi^2\Gamma_\phi^2}{4m_\phi\Gamma_\phi}
    \arctan\left(\frac{\zeta_+ - \zeta_-}{1 + \zeta_+\zeta_-}\right)\right.
    \nonumber\\ 
    &\hspace{12em} \left.+ \frac{1}{4}(m_\phi^2 - m_{\gamma'}^2)
    \log\left(\frac{\zeta_+^2 + 1}{\zeta_-^2 + 1}\right)\right\},
\label{eq:integrals-gg}
\end{align}
for $f = \gamma\gamma'$ and
\begin{align}
    I_{e^+e^-}(T) &= 
    \int \left[\prod_{i = 3}^4 \frac{d^3\vec{p}_i}{(2\pi)^3 2 E_i}\right] 
    \frac{(p_3 \cdot p_4)+m_e^2}{(p_\phi^2 - m_\phi^2)^2 + m_\phi^2 \Gamma_\phi^2} 
    \delta(Q - T - E_3 - E_4) \nonumber\\
    &= \frac{1}{4(2\pi)^4} \int_{m_e}^{Q - T - m_e} dE_3 
    \left\{\frac{m_\phi}{2\Gamma_\phi}
    \arctan\left(\frac{\zeta_+ - \zeta_-}{1 + \zeta_+\zeta_-}\right) 
    + \frac{1}{4}\log\left(\frac{\zeta_+^2 + 1}{\zeta_-^2 + 1}\right)\right\},
\label{eq:integrals-ee}
\end{align}
for $f = e^+e^-$. Here, we define $\zeta_\pm = 2|\vec{p}_3||\vec{p}_4|(x \pm 1)/(m_\phi \Gamma_\phi)$ and
\begin{align}
    x = \frac{2E_3 E_4 + m_3^2 + m_4^2 - m_\phi^2}{2|\vec{p}_3||\vec{p}_4|},
\end{align}
with $|\vec{p}_i| = \sqrt{E_i^2 - m_i^2}$ ($i=3,4$), $E_4 = Q - T - E_3$. The quantity $x$ corresponds to the cosine of the opening angle $\theta_{34}$ between the $\phi$ decay product momenta when $\phi$ is on-shell, as explored later in this appendix.

The remaining phase space of the prompt electrons can be reduced to the total electron kinetic energy, $T = E_{e_1}^\text{kin} + E_{e_2}^\text{kin}$, by integrating over the energy difference $\Delta T = E_{e_1}^\text{kin} - E_{e_2}^\text{kin}$ and the angle between $\vec{p}_1$ and $\vec{p}_2$. We then obtain the differential rates
\begin{align}
\label{eq:rates}
    \frac{d\Gamma_{\nu\nu}}{dT} &= 
    \kappa_{0\nu}\left(\frac{m_e}{2R}\right)^2
    \left|\mathcal{M}_{0\nu}\right|^2 c_\nu^4 I_{\nu\nu}(T)g(T)\Theta(0 < T < Q), \nonumber\\
    \frac{d\Gamma_{\gamma\gamma'}}{dT} &= 
    \frac{\kappa_{0\nu}}{1+\delta_{\gamma\gamma'}}
    \left(\frac{m_e}{2R}\right)^2
    \left|\mathcal{M}_{0\nu}\right|^2 c_\nu^2 g_{\phi\gamma\gamma'}^2 
    I_{\gamma\gamma'}(T)g(T)\Theta(0 < T < Q - m_{\gamma'}), \nonumber\\
    \frac{d\Gamma_{e^+e^-}}{dT} &= 
    2\kappa_{0\nu}\left(\frac{m_e}{2R}\right)^2
    \left|\mathcal{M}_{0\nu}\right|^2 c_\nu^2 c_e^2 I_{e^+e^-}(T)g(T)\Theta(0 < T < Q - 2m_e),
\end{align}
with the pre-factor $\kappa_{0\nu} = (\mathcal{N}_{0\nu} R)^2/(2\pi^4 m_e^2) = G_F^4 \cos^4\theta_C/(8\pi^5 m_e^2)$ and the double-sided Heaviside function
\begin{align}
    \Theta(a < x < b) \equiv \Theta(x - a) \Theta(b - x).
\end{align}
The integral over the electron kinetic energy difference $\Delta T$ is captured by the function
\begin{align}
\label{eq:g(T)}
    g(T) = 
    \int_{-T}^T d\Delta T |\vec{p}_1||\vec{p}_2| E_1 E_2 F^2(E_1, E_2),
\end{align}
where $E_1 = \frac{1}{2}(T+\Delta T) + m_e$ and $E_2 = \frac{1}{2}(T-\Delta T) + m_e$. We note that if the $\phi \to e^+e^-$ decay occurs a distance at which the Fermi functions cannot be neglected, the factorisation of the phase space above can no longer be performed.

\subsection{Effective operators}

It is useful to explore different limits of Eq.~\eqref{eq:rates}. In the limit $m_\phi\gg Q$, $\phi$ can effectively be integrated out of the theory to yield the contact effective operators at dimension $d\leq 7$
\begin{align}
\label{eq:L_EFT}
    \mathcal{L} &\supset 
    \frac{c_{\nu\nu}^{ijkl}}{8}(\bar{\nu}_i P_L \nu_j)(\bar{\nu}_k P_L \nu_l) 
    + \frac{c_{\nu\nu}^{\prime ijkl}}{8}(\bar{\nu}_i P_L \nu_j)(\bar{\nu}_k P_R \nu_l)  
    \nonumber \\
    &+ \frac{c_{\nu e}^{ijkl}}{2}(\bar{\nu}_i P_L \nu_j)(\bar{e}_k P_L e_l) + \frac{c_{\nu e}^{\prime ijkl}}{2}(\bar{\nu}_i P_L \nu_j)(\bar{e}_k P_R e_l) \nonumber\\
    &+ \frac{c_{e e}^{ijkl}}{2}(\bar{e}_i P_L e_j)(\bar{e}_k P_L e_l) 
    + \frac{c_{e e}^{\prime ijkl}}{2}(\bar{e}_i P_L e_j)(\bar{e}_k P_R e_l) 
    \nonumber\\
    &+ \frac{c_{\nu \gamma\gamma}^{ij}}{8}(\bar{\nu}_i P_L \nu_j)F_{\mu\nu}\tilde{F}^{\mu\nu} 
    + \frac{c_{\nu \gamma \gamma_D}^{ij}}{4}(\bar{\nu}_i P_L \nu_j)F_{\mu\nu}\tilde{F}_D^{\mu\nu}  \nonumber\\
    &+ \frac{c_{e \gamma\gamma}^{ij}}{4}(\bar{e}_i P_L e_j)F_{\mu\nu}\tilde{F}^{\mu\nu}  + \frac{c_{e \gamma \gamma_D}^{ij}}{2}(\bar{e}_i P_L e_j)F_{\mu\nu}\tilde{F}_D^{\mu\nu} 
    + \text{h.c.},
\end{align}
with the effective Wilson coefficients
\begin{align}
\label{eq:matching}
    c_{ff'}^{ijkl} = \frac{c_{f}^{ij} c_{f'}^{kl}}{m_\phi^2}, \quad c_{ff'}^{\prime ijkl} = \frac{c_{f}^{ij} c_{f'}^{kl*}}{m_\phi^2}, \quad
    c_{f \gamma \gamma'}^{ij} = \frac{c_{f}^{ij} g_{\phi\gamma\gamma'}}{m_\phi^2},
\end{align}
with $f, f' = \nu, e$. In Eq.~\eqref{eq:L_EFT}, the first two $d = 6$ operators are neutrino self-interactions ($\nu$SI) which mediate the $2\nu\beta\beta$ process, as studied for instance in Ref.~\cite{Deppisch:2020sqh}. The first (second) $\nu$SI operator is equivalent to the lepton number violating (conserving) operator in Ref.~\cite{Deppisch:2020sqh}. We also obtain the $d = 7$ Rayleigh operators coupling neutrinos to gauge bosons (or \textit{neutrino polarisability}, generalized to include couplings to dark photons). As we consider $\phi$ only coupling to the first generation leptons, i.e., $c_\nu^{ij} =  i U_{ei} U_{ej} c_\nu$ and $c_e^{ij} = i\delta_{ei}\delta_{ej} c_e$, we can define the effective coefficients $c_{\nu\nu} \equiv c_\nu^2/m_\phi^2$, $c_{\nu\nu}' \equiv |c_\nu|^2/m_\phi^2$, $c_{\nu e} \equiv c_\nu c_e/m_\phi^2$, $c_{\nu e}' \equiv c_\nu c_e^*/m_\phi^2$, and $c_{\nu\gamma\gamma'} = c_\nu g_{\phi\gamma\gamma'}/m_\phi^2$.

In the heavy $\phi$ limit, the phase space integrals in Eqs.~\eqref{eq:integrals-nunu} and~\eqref{eq:integrals-gg} can be written as $I_f(T) = \hat{I}_f(T)/m_\phi^4$, with
\begin{align}
    \hat{I}_{\nu\nu}(T) = \frac{(Q - T)^5}{30(2\pi)^4},
\end{align}
for $f = \nu\nu$,
\begin{align}
    \hat{I}_{\gamma\gamma'}(T) &= \frac{(8(Q-T)^6  - 38 (Q-T)^4 m_{\gamma'}^2 + 87 (Q-T)^2 m_{\gamma'}^4 + 48m_{\gamma'}^6\big)\sqrt{(Q-T)^2 - m_{\gamma'}^2}}{840(2\pi)^4} \nonumber \\
    &\hspace{1.3em} + \frac{(Q-T)m_{\gamma'}^6}{8(2\pi)^4}\ln \frac{m_{\gamma'}}{Q- T + \sqrt{(Q-T)^2 - m_{\gamma'}^2}}
    \nonumber\\
    &= \frac{(Q - T)^7}{105(2\pi)^4} \qquad (m_{\gamma'} = 0),
\end{align}
for $f = \gamma\gamma'$. The differential rates for assuming an electron flavor neutrino coupling are then given in terms of the EFT coefficients defined below Eq.~\eqref{eq:matching}, as
\begin{align}
\label{eq:rates_EFT}
    \frac{d\Gamma_{\nu\nu}}{dT} &= 
    \frac{\kappa_{0\nu}}{2}\left(\frac{m_e}{2R}\right)^2 \left|\mathcal{M}_{0\nu}\right|^2 \left(|c_{\nu\nu}|^2 +|c_{\nu\nu}'|^2\right) \hat{I}_{\nu\nu}(T) g(T) \Theta(0 < T < Q), \nonumber\\
    \frac{d\Gamma_{\gamma\gamma'}}{dT} &= 
    \frac{\kappa_{0\nu}}{1+\delta_{\gamma\gamma'}}
    \left(\frac{m_e}{2R}\right)^2 \left|\mathcal{M}_{0\nu}\right|^2 
    \left|c_{\nu\gamma\gamma'}\right|^2 
    \hat{I}_{\gamma\gamma'}(T) g(T) \Theta(0 < T < Q - m_{\gamma'}).
\end{align}
As mentioned in the previous section, the presence of the Fermi functions for prompt emitted $e^+e^-$ prevents the factorization of the phase space with respect to the prompt electrons. Thus, we do not give an equivalent expression for $f = e^+e^-$ in the heavy $\phi$ limit.

\subsection{On-shell production}
\label{subsec:on-shell}

In the limit $m_\phi < Q$, $\phi$ can be produced on-shell for electron energies $T < Q - m_\phi$. While off-shell production is also possible in the regime $T > Q - m_\phi$, it is highly suppressed for small couplings and therefore has a small decay width $\Gamma_\phi$. The relevant expressions are significantly simplified if one factorizes the phase space in Eq.~\eqref{eq:dlips} by inserting the identity $1 = \int d^4 p_\phi \delta^{(4)}(p_\phi - p_3 - p_4)$ with $d^4 p_\phi = dE_\phi d^3\vec{p}_\phi = dp_\phi^2 d^3 \vec{p}_\phi/(2E_\phi)$,
\begin{align}
    d\Pi_4 &= 
    \left[\prod_{i = 1}^2 \frac{d^3\vec{p}_i}{(2\pi)^3 2 E_i}\right] 
    \frac{d^3\vec{p}_\phi}{(2\pi)^3 2 E_\phi} 
    \delta\left(Q - T - E_\phi\right)
    \times \frac{1}{2\pi}dp_\phi^2 \times d\Pi_\text{decay},
\end{align}
with the decay phase space element
\begin{align}
    d\Pi_\text{decay} = 
    \left[\prod_{i = 3}^4 \frac{d^3\vec{p}_i}{(2\pi)^3 2 E_i}\right]
    (2\pi)^4\delta^{(4)}(p_\phi - p_3 - p_4).
\end{align}
Here, $p_\phi$ is generally off-shell with $p_\phi^2 \neq m_\phi^2$ and the energy determined as $E_\phi^2 = p_\phi^2 + |\vec{p}_\phi|^2$. Using the narrow-width approximation (NWA) for the $\phi$ propagator, e.g., in Eq.~\eqref{eq:phase_space_factors-gg},
\begin{align}
    \frac{1}{(p_\phi^2 - m_\phi^2)^2 + m_\phi^2 \Gamma_\phi^2} \xrightarrow{\Gamma_\phi \ll m_\phi}
    \frac{\pi}{m_\phi \Gamma_\phi} \delta(p_\phi^2 - m_\phi^2) ,
\end{align}
factorizes the calculation into the production of $\phi$ from massive Majoron-like double beta decay ($0\nu\beta\beta\phi$) and decay $\phi \to f$, by setting $\phi$ on-shell with $p_\phi^2 = m_\phi^2$. Due to the Lorentz-scalar nature of $\phi$, it is emitted isotropically from the nucleus and is uncorrelated with the prompt electron momenta directions. The $\phi$ phase space integration can therefore be expressed as
\begin{align}
    \int_{\Omega_\phi}\frac{d^3\vec{p}_\phi}{(2\pi)^3 2 E_\phi} =
    \frac{1}{4\pi^2}\sqrt{(Q - T)^2 - m_\phi^2} dT,
\end{align}
where $T = Q - E_\phi$ is the usual kinetic energy release in the double beta decay electrons due to overall energy conservation. Thus, the phase space decomposes as 
\begin{align}
\label{eq:dGfull}
    d\Gamma_{ij} &= 
    \frac{d\Gamma_\phi}{dT} dT
    \times \frac{\text{Br}(\phi\to f)}{4\pi} 
    \delta\left(E_3' - E_3^*\right) 
    dE_3' d\cos\theta_3' d\phi_3',
\end{align}
where the phase space of the decay is expressed in the rest frame of $\phi$, where the energy and momentum of one of the $\phi$ decay products is
\begin{align}
E_3^* = \frac{m_\phi^2 + m_3^2 - m_4^2}{2m_{\phi}},\quad |\vec{p}_3^*| = \frac{\lambda^{1/2}(m_\phi^2, m_3^2, m_4^2)}{2m_\phi} ,
\label{eq:phi_rest_frame}
\end{align}
respectively, where $\lambda(x,y,z) = x^2 + y^2 + z^2 - 2(x y + y z + z x)$ is the Källén function. Here, the usual rate for the production of a massive pseudoscalar is
\begin{align}
\label{eq:Majoron-rate}
    \frac{d\Gamma_\phi}{dT} &= 
    \frac{\kappa_{0\nu}}{8\pi^2}\left(\frac{m_e}{2R}\right)^2 
    \left|\mathcal{M}_{0\nu}\right|^2 c_\nu^2 g(T)
    \sqrt{(Q - T)^2 - m_\phi^2} 
    \Theta(0 < T < Q - m_\phi),
\end{align}
and the branching ratios for $\phi\to\nu\nu$, $\phi\to\gamma\gamma'$ and $\phi\to e^+e^-$, respectively, are
\begin{align}
    \text{Br}(\phi\to\nu\nu) &= 
        \frac{c_\nu^2 m_\phi}{16\pi \Gamma_\phi}, \\
    \text{Br}(\phi\to\gamma\gamma') &= 
        \frac{g_{\phi\gamma\gamma'}^2 m_\phi^3}{32\pi(1 + \delta_{\gamma\gamma'})\Gamma_\phi}
        \left(1 - \frac{m_{\gamma'}^2}{m_\phi^2}\right)^3, \\
    \text{Br}(\phi\to e^+e^-) &= 
        \frac{c_e^2 m_\phi}{8\pi\Gamma_\phi}
        \left(1 - \frac{4m_e^2}{m_\phi^2}\right)^{1/2}.
\end{align}
\begin{figure}[t!]
    \centering
    \includegraphics[width=0.6\textwidth]{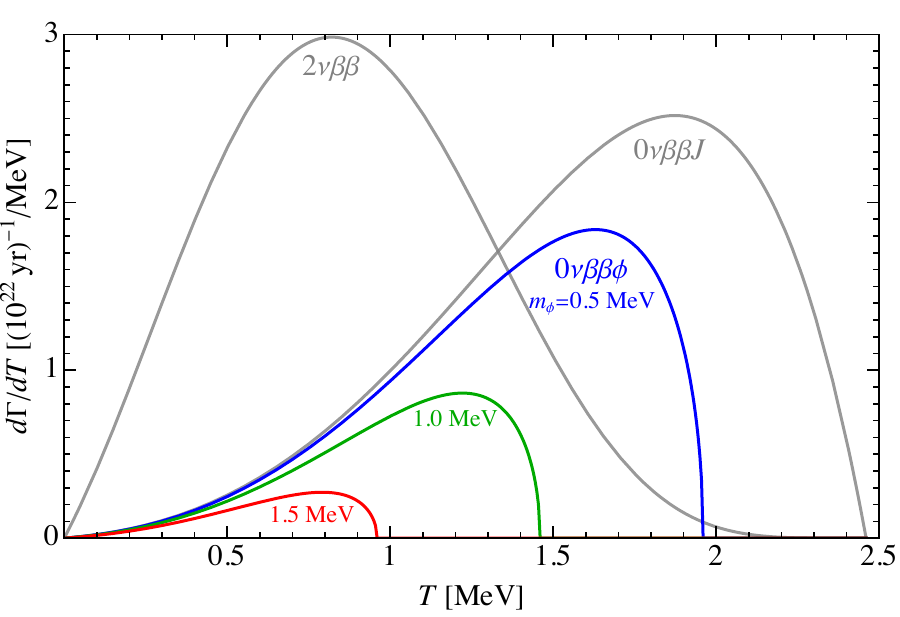}
    \caption{Spectrum of $0\nu\beta\beta\phi$ decay for $^{136}$Xe with respect to the prompt electron kinetic energy $T$ for three different pseudoscalar masses $m_\phi = 0.5$, 1, 1.5~MeV and the coupling $c_\nu = 3.5\times 10^{-4}$. Also shown are the spectra for standard (massless) Majoron emission $0\nu\beta\beta J$, for the same $c_\nu$ coupling, and SM $2\nu\beta\beta$ decay.}
    \label{fig:dGdT}
\end{figure}
If $\phi$ is stable over the size of the detector, or if it decays to invisible states only, Eq.~\eqref{eq:Majoron-rate} gives the observable decay distribution with respect to the kinetic energy release in the electrons. It is shown in Fig.~\ref{fig:dGdT} for $^{136}$Xe and three different $\phi$ masses. For comparison, we also show the spectrum for ordinary, i.e., massless Majoron emission $0\nu\beta\beta J$, equivalent to Eq.~\eqref{eq:Majoron-rate} with $m_\phi = 0$, and the SM $2\nu\beta\beta$ spectrum,
\begin{align}
\label{eq:2nbb-SM}
    \frac{d\Gamma^{2\nu}_\text{SM}}{dT} = 
    \frac{\kappa_{0 \nu}}{2 \pi^2} |\mathcal{M}_{2\nu}|^2 g(T) \frac{1}{30}\left(Q-T\right)^5
    \Theta(0 < T < Q),
\end{align}
neglecting any corrections due to the final state momentum dependence of the $2\nu\beta\beta$ NME $\mathcal{M}_{2\nu}$ \cite{Boudjema:2025okq}.

If it has a sufficiently small width, $\phi$ decays at a macroscopic distance from where it was emitted in the $0\nu\beta\beta\phi$ decay process. If $\phi$ decays to (practically) invisible particles such as two neutrinos or dark photons, this cannot be observed, but one or two photons or an $e^+e^-$ pair gives rise to potentially displaced vertices that can be detected in double beta decay experiments. In this case, the location of the displaced vertex (or the time delay), in relation to the primary double beta decay, and the energy released can be detected. Mathematically, the spatial coordinate part of the phase space, implicitly integrated over above, has to be considered. In this case, the decay length $L$ follows the normalized distribution
\begin{align}
    p(L) dL = 
    \frac{1}{\gamma\beta L_\phi} \exp\left(-\frac{L}{\gamma\beta L_\phi}\right) \Theta(0 < L) dL.
\end{align}
where $L = |\mathbf{x}_D - \mathbf{x}_P|$ is the distance between the primary and displaced vertex and $L_\phi = 1 / \Gamma_\phi$ is the proper decay length of $\phi$. The relativistic factor $\gamma\beta$ can be expressed in terms of $E_\phi = Q - T$ as $\gamma\beta = |\vec{p}_\phi|/m_\phi = \sqrt{(E_\phi/m_\phi)^2 - 1}$.

By applying a Lorentz transformation from the $\phi$ rest frame to the laboratory frame, where $\phi$ is boosted with the velocity $\beta = |\vec{p}_\phi|/E_\phi = \sqrt{1 - (m_\phi/E_\phi)^2}$ along the direction $(\mathbf{x}_D - \mathbf{x}_P)/L$, the decay phase space element in Eq.~\eqref{eq:dGfull} transforms as
\begin{align}
    &\frac{1}{4\pi} \delta\left(E_3' - E_3^*\right) 
    dE_3' d\cos\theta_3' d\phi_3' 
    \\ \nonumber
    &\to 
    \frac{|\mathcal{J}|}{4\pi}\delta\left(\gamma (E_3 - \beta|\vec{p}_3|\cos\theta_3) - E_3^*\right) 
     dE_3 d\cos\theta_3 d\phi_3,
     \label{eq:photon_phasespace_boost}
\end{align}
with $\gamma = E_\phi/m_\phi$, where primed (unprimed) variables denote the energy and angles of the decay product in the $\phi$ rest frame (laboratory frame). The angle $\theta_3$ is with respect to the decay product direction and the displacement vector, $\cos\theta_3 = \vec{p}_3 \cdot (\mathbf{x}_D - \mathbf{x}_P)/|\vec{p}_3|L$, and $\phi_3$ is the azimuthal angle around the displacement vector. The Jacobian of the boost is
\begin{align}
\mathcal{J} = \frac{|\vec{p}_3|}{\sqrt{\gamma^2(E_3 - \beta |\vec{p}_3|\cos\theta_3)^2 - m_3^2}}.
\end{align}

The laboratory frame energy $E_3$ and angle $\theta_3$ are given in terms of the rest frame angle as
\begin{align}
\label{eq:boost_relations}
E_3 = \gamma(E_3^* + \beta |\vec{p}_3^*|\cos\theta_3') ,\quad \cos\theta_3 = \frac{\gamma(\beta E_3^* + |\vec{p}_3^*|\cos\theta_3')}{\sqrt{\gamma^2(E_3^* + \beta |\vec{p}_3^*| \cos\theta_3')^2 - m_3^2}},
\end{align}
respectively, with $E_3^*$ and $|\vec{p}_3^*|$ given in Eq.~\eqref{eq:phi_rest_frame}. Two values of $\theta_3'$ can yield the same value of $\theta_3$ when $\beta > |\vec{p}_3^*|/E_3^*$, i.e. when the $\phi$ velocity in the laboratory frame is greater than the velocity of the decay product in the $\phi$ rest frame. In terms of the energy of $\phi$, this condition is equivalent to $E_\phi > m_\phi E_3^*/m_3$. In this limit, there is a minimum value of $\cos\theta_3$ at the value $\cos\theta_3' = - |\vec{p}_3^*|/\beta E_3^*$, where $E_3 = \gamma m_3^2/E_3^*$. We may write the minimum $\cos\theta_3$ as
\begin{align}
\cos\theta_3^{-} = 
\begin{cases}
-1 & E_\phi < m_\phi E_3^*/m_3 \\
\frac{E_3^*}{m_3}\sqrt{1 - \frac{|\vec{p}_3^*|^2}{\beta^2 (E_3^*)^2}} & E_\phi \geq m_\phi E_3^*/m_3
\end{cases}.
\label{eq:minimum_theta3}
\end{align}
Thus, while the angles in the $\phi$ rest frame satisfy $-1 \leq \cos\theta_3' \leq 1$ and $0 < \phi_3' < 2\pi$, in the laboratory frame they have $\cos\theta_3^- \leq \cos\theta_3 \leq 1$ and $0 < \phi_3 < 2\pi$. 
The expressions above simplify for $m_3 = 0$ (relevant for $\gamma\gamma'\beta\beta$ decay), with the Jacobian given by $\mathcal{J} = 1/\gamma(1 - \beta\cos\theta_3)$ and $\theta_3$ being a monotonic function of $\theta_3'$ with $-1 \leq \cos\theta_3 \leq 1$. For $m_3 = 0$, the energy and angle in the laboratory frame follow the simple relation,
\begin{align}
\label{eq:boost_relation}
    E_3 = \frac{E_3^*}{\gamma(1 - \beta\cos\theta_3)} = 
    \frac{m_\phi E_3^*}{E_\phi - |\vec{p}_\phi|\cos\theta_3}.
\end{align}
Note that all of the expressions above apply to the other final state particle, with $3 \to 4$.

Thus, we can determine the fully differential distribution with respect to the (in principle) observable electron energy release $T$, final state particle energy $E_3$ and angle $\cos\theta_3$, and displacement $L$ as
\begin{align}
    \frac{d\Gamma}{dT dE_3 d\cos\theta_3 dL} &=  
    \frac{\kappa_{0\nu}}{16\pi^2}\left(\frac{m_e}{2R}\right)^2 
    \left|\mathcal{M}_{0\nu}\right|^2 c_\nu^2 
    \frac{m_\phi}{L_\phi}\text{Br}(\phi\to f) \nonumber\\  
    &\times
    \frac{|\vec{p}_3|g(T)}{\sqrt{\gamma^2(E_3 - \beta |\vec{p}_3|\cos\theta_3)^2 - m_3^2}} 
    \exp\left[-\frac{L}{\gamma\beta L_\phi}\right]
    \nonumber\\
    &\times \Theta(0 < T < Q - m_\phi)\Theta(0 < L)\nonumber\\
    &\times \Theta(E_3^- < E_3 < E_3^+) \Theta(-1 < \cos\theta_3 < 1) \nonumber\\
    &\times 
    \delta\left(\gamma (E_3 - \beta|\vec{p}_3|\cos\theta_3) - E_3^*\right),
\label{eq:distro-Egamma-costheta3-L}
\end{align}
where the final state energy $E_3$ in the laboratory frame has the maximal and minimal values at $T = 0$, i.e.
\begin{align}
\label{eq:boost_relation-limits}
    E_3^\pm = 
    \frac{E_3^*Q \pm |\vec{p}_3^*|\sqrt{Q^2 - m_\phi^2}}{m_\phi},
\end{align}
respectively. Due to the Dirac delta function in Eq.~\eqref{eq:distro-Egamma-costheta3-L},
the four kinematic variables are not independent of each other, but the above expression can be integrated in various ways to yield experimentally relevant distributions as discussed below. 

\begin{figure}[t!]
    \centering
    \includegraphics[width=0.49\linewidth]{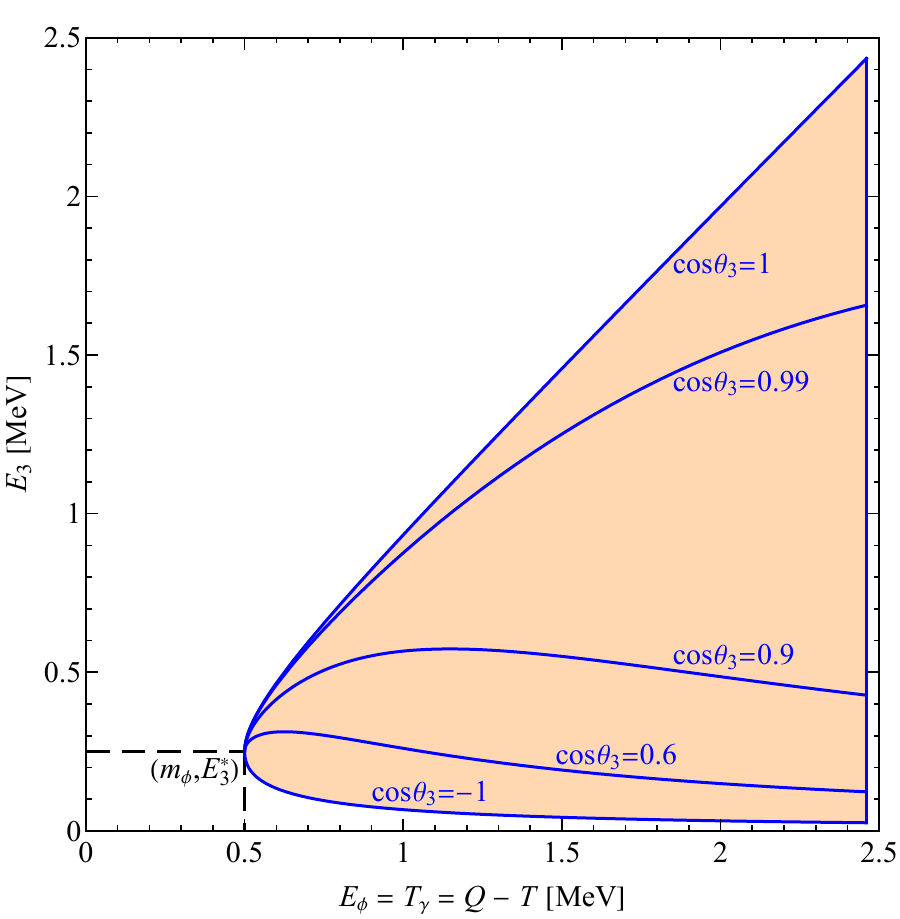}
    \includegraphics[width=0.49\linewidth]{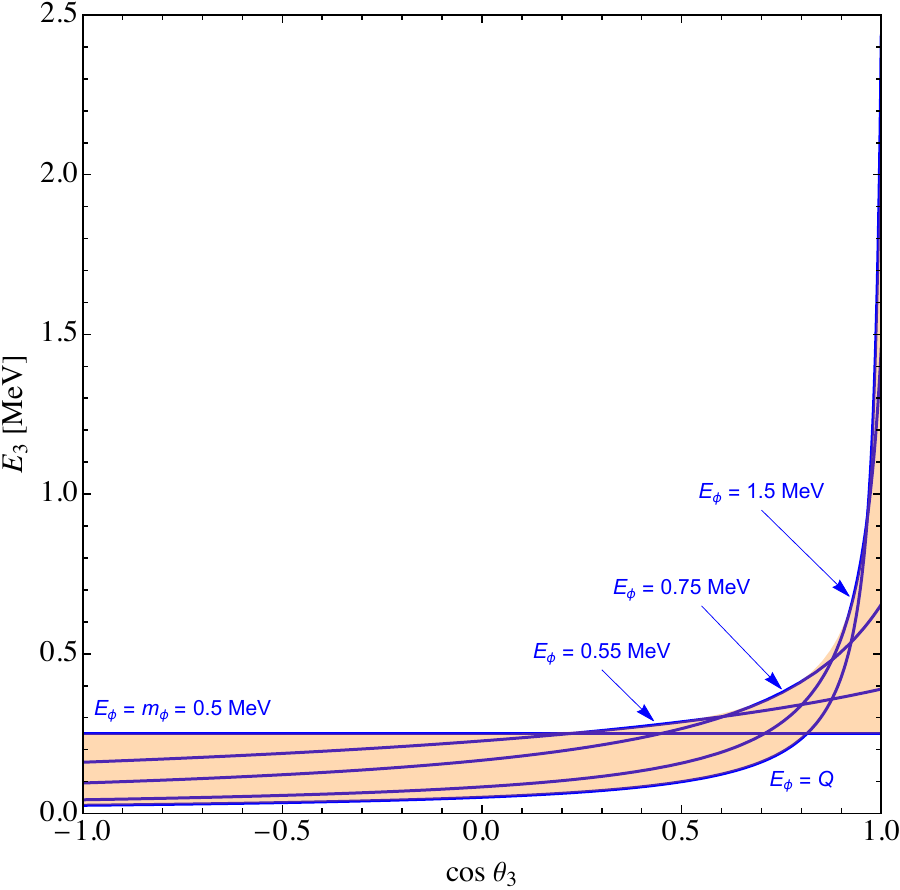}
    \caption{For $\gamma\gamma\beta\beta$ decay with $m_\phi = 0.5$~MeV, the kinematic relations in the laboratory frame between $E_3$ and $E_\phi$ (left) for different $\cos\theta_3$ values and $E_3$ and $\cos\theta_3$ (right) for different $E_\phi$ values. The orange shading indicates the physically allowed region.}
    \label{fig:kinematic}
\end{figure}

For $m_3 = m_4 = 0$ (relevant for $\gamma\gamma\beta\beta$ decay), the relation between the kinematic variables in the laboratory frame, namely $E_3$ as a function of $E_\phi = Q - T$ and $\cos\theta_3$ in Eq.~\eqref{eq:boost_relation}, is displayed in Fig.~\ref{fig:kinematic}. The shaded region denotes the kinematically allowed regime. The range in $E_\phi = E_3 + E_4$ is $m_\phi < E_\phi < Q$. For a given $E_\phi$, the maximal and minimal values $E_3^{\pm}(E_\phi)$ are given by  Eq.~\eqref{eq:boost_relation} for $\cos\theta_3 = \pm 1$, i.e., for the photon in the forward and backward directions, respectively. The overall extreme $E_3$ values are achieved for the highest $\phi$ boost, i.e., when $E_\phi = Q$, see Eq.~\eqref{eq:boost_relation-limits}. For $E_\phi = m_\phi$, the photon energy approaches its unique value in the $\phi$ rest frame, $E_3^{\pm} = E_3^*$. 
The outgoing angle is also in the range $-1 < \cos\theta_3 < 1$ in every frame. The phase space depicted in Fig.~\ref{fig:kinematic} is dominated by angles in the forward direction, $\cos\theta_3 > 0$. Note that for a given $E_3$, there are generally two solutions for $E_\phi$ if $\cos\theta_3 > 0$ (see curve for $\cos\theta_3 = 0.9$), though one solution may occur in the kinematically inaccessible regime $E_\phi > Q$.

Along with analytical expressions, in the following subsections we illustrate the decay distributions for the $\gamma\gamma\beta\beta$ and $\gamma\gamma_D\beta\beta$ decay scenarios. For these, we choose the benchmark couplings
\begin{align}
\label{eq:benchmark}
    c_\nu = 2.23 \times 10^{-6}, \quad 
    g_{\phi\gamma\gamma'} = 6.31 \times 10^{-6}
    \sqrt{1+\delta_{\gamma \gamma^\prime}},
\end{align}
which correspond to a $\phi$ width and proper decay length,
\begin{align}
    L_\phi = 2~\text{m} 
    = \Gamma_\phi^{-1} 
    = \left(9.9 \times 10^{-14} ~\text{MeV}\right)^{-1} 
    \quad \text{for} \quad
    m_\phi = 0.5~\text{MeV},
\end{align}
and equal branching ratios, $\text{Br}(\phi\to\gamma\gamma') = \text{Br}(\phi\to\nu\nu) = 1/2$.

\subsubsection{Fully visible final state}

Here we discuss the case where $\phi$ decays to two visible SM states, $\phi\to\gamma\gamma$ or $\phi \to e^+e^-$, and where the detector is sensitive to the total energy $T_f = E_3 + E_4$ (in the following, we use $T_f = T_\gamma$ for $\gamma\gamma\beta\beta$ decay and $T_f = T_e$ for $e^+e^-\beta\beta$ decay) released in the displaced vertex. From energy conservation, $T_f = E_\phi = Q - T$. It is then straightforward to transform $T \to T_f$ in Eq.~\eqref{eq:distro-Egamma-costheta3-L}.

\paragraph*{Distribution with respect to $T_f$ and $L$.} Integrating over $E_3$ and $\cos\theta_3$ in Eq.~\eqref{eq:distro-Egamma-costheta3-L}, the double differential decay distribution with respect to the total $\gamma\gamma$ or $e^+e^-$ energy $T_f$ and the $\phi$ displacement $L$ is
\begin{align}
\label{eq:distro-Tgamma-L}
    \frac{d\Gamma}{dT_f dL} &=  
    \frac{\kappa_{0\nu}}{8\pi^2}\left(\frac{m_e}{2R}\right)^2 
    \left|\mathcal{M}_{0\nu}\right|^2 c_\nu^2 
    \frac{m_\phi}{L_\phi} \text{Br}(\phi\to f) \nonumber\\  
    &\times g(Q - T_f) 
    \exp\left[-\left(\frac{T_f^2}{m_\phi^2} - 1\right)^{-1/2}\frac{L}{L_\phi}\right] 
    \Theta(m_\phi < T_f < Q) \Theta(0 < L).
\end{align}
\begin{figure}[t!]
    \centering
    \includegraphics[width=0.6\linewidth]{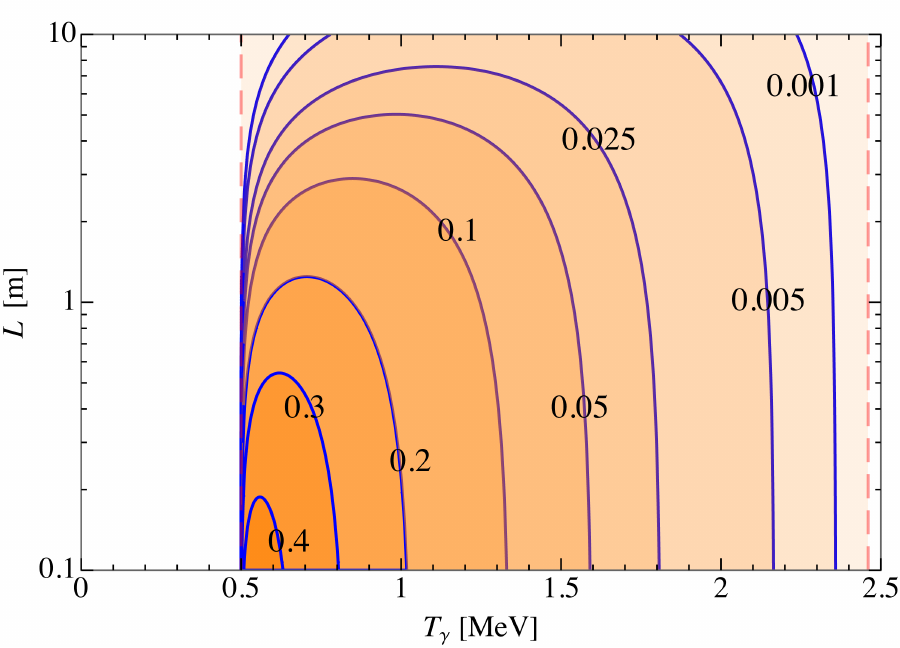}
    \caption{Normalized $\gamma\gamma\beta\beta$ decay distribution $\Gamma^{-1}d\Gamma/(dT_\gamma dL)$ for $^{136}$Xe with respect to the total photon energy $T_\gamma$ and $\phi$ displacement $L$ with $m_\phi = 0.5$~MeV and benchmark couplings in Eq.~\eqref{eq:benchmark}.}
    \label{fig:2D_Tgamma_L}
\end{figure}
The normalized distribution $\Gamma^{-1}d\Gamma/(dT_\gamma dL)$ is shown in Fig.~\ref{fig:2D_Tgamma_L} for the $\gamma\gamma\beta\beta$ decay of $^{136}$Xe, $m_\phi = 0.5$~MeV and the benchmark couplings in Eq.~\eqref{eq:benchmark}, corresponding to a proper decay length $L_\phi = 2$~m. As intuitively expected, there is a correlation between the di-photon energy and the $\phi$ displacement: the larger the energy, the larger the $\phi$ boost and the longer the displacement on average. This effect will be reduced for heavier $\phi$ with a more restricted range in boost. Due to the on-shell production of $\phi$, the photons will have energies $T_\gamma > m_\phi$.

\begin{figure}[t!]
    \centering
    \includegraphics[width=0.6\textwidth]{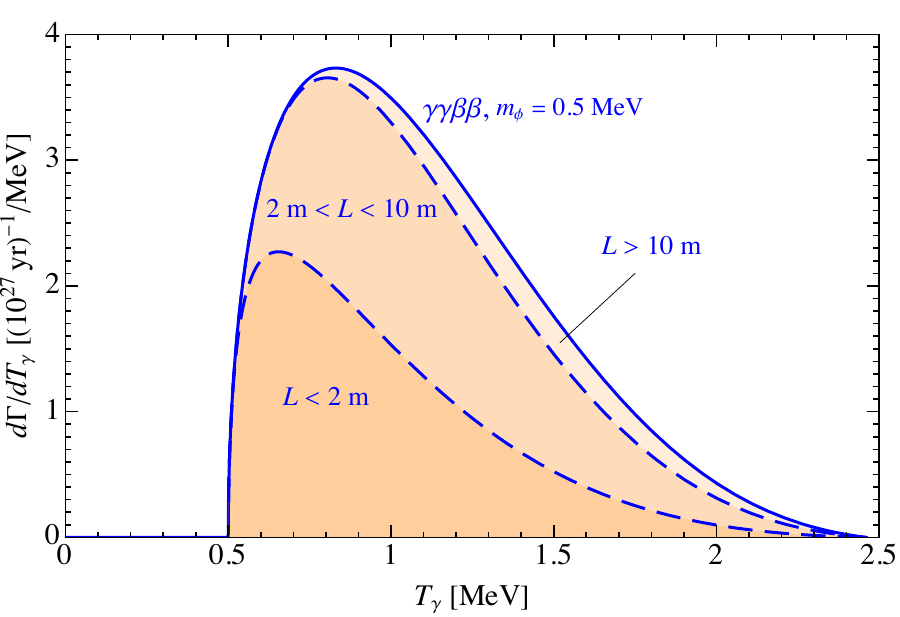}
    \caption{Distribution of $\gamma\gamma\beta\beta$ decay for $^{136}$Xe with respect to the total photon energy $T_\gamma$ for $m_\phi = 0.5$~MeV and the benchmark couplings in Eq.~\eqref{eq:benchmark}. The shaded regions indicate the contributions where $\phi$ decays at distances $L < 2$~m, $2~\text{m} < L < 10$~m and $L > 10$~m as indicated.}
    \label{fig:dGdTgamma}
\end{figure}

\paragraph*{Distribution with respect to $T_f$.} The spectrum with respect to $T_f$ only, but for decays within the distance range $L_1 < L < L_2$, can be determined from Eq.~\eqref{eq:distro-Tgamma-L} by integrating analytically over $L$,
\begin{align}
\label{eq:distro-Tgamma}
    \left.\frac{d\Gamma}{dT_f}\right|_{L_1}^{L_2} &=  
    \frac{\kappa_{0\nu}}{8\pi^2}\left(\frac{m_e}{2R}\right)^2 
    \left|\mathcal{M}_{0\nu}\right|^2 c_\nu^2 
    \text{Br}(\phi\to f) \nonumber\\  
    &\times g(Q - T_f) \sqrt{T_f^2 - m_\phi^2} 
    \exp\left[-\left(\frac{T_f^2}{m_\phi^2} - 1\right)^{-1/2}\frac{L}{L_\phi}\right]_{L_2}^{L_1}\Theta(m_\phi < T_f < Q)    .
\end{align}
This distribution is expected to be experimentally relevant if the $\gamma\gamma$ or $e^+e^-$ energy releases can be measured within certain distance ranges of the detector. This is shown in Fig.~\ref{fig:dGdTgamma} in the benchmark $\gamma\gamma\beta\beta$ decay scenario considered. The shaded regions indicate the contributions to the spectrum $d\Gamma/dT_\gamma$ from decays in three distance ranges, $L < 2$~m, $2~\text{m} < L < 10$~m and $L > 10$~m. In the given scenario with $L_\phi = 2$~m, around 46\% of decays occur between 2~m and 10~m and, as mentioned above, the produced photons have, on average, a higher energy than in more rapid decays.

Integrating Eq.~\eqref{eq:distro-Tgamma} over the photon energy gives the total $\gamma\gamma\beta\beta$ or $e^+e^-\beta\beta$ decay rate of fully visible $\gamma\gamma$ or $e^+e^-$ within a distance range $L_1 < L < L_2$,
\begin{align}
\label{eq:totGamma}
    \Gamma(L_1 < L < L_2) &=
    \int_{m_\phi}^Q dT_f 
    \left.\frac{d\Gamma}{dT_f}\right|_{L_1}^{L_2}.
\end{align}

\paragraph*{Distribution with respect to $T_\text{tot}$.}

As all of the double beta decay energy $Q$ is released to visible particles, the corresponding distribution with respect to the total energy $T_\text{tot} = T + T_f$ is a Dirac delta function at the double beta decay endpoint $Q$, as long as the $\phi$ decay occurs within the detector. For detectors that cannot distinguish between the electron and photon energy release, at least to a minimal distance $L_\text{min}$ where a secondary vertex cannot be resolved, this is the main signature in this case. This signal is then identical to standard $0\nu\beta\beta$ decay, benefiting from the optimized low background at the endpoint. If $\phi$ decays outside the detector, the contribution is equivalent to the continuous spectrum of massive Majoron emission with respect to $T$ only, as discussed above. The latter also receives contributions from invisible decays, i.e., $\phi\to\nu\nu$, at any distance. This gives a combined spectrum of
\begin{align}
\label{eq:distro_Ttotal_full}
    \left.\frac{d\Gamma}{dT_\text{tot}}\right|_0^{L_\text{min}} &=
    \Gamma(L < L_\text{min}) \times \delta(T_\text{tot} - Q) \nonumber \\
    &\hspace{-1em}+ \frac{d\Gamma_\phi}{dT}\left\{
    \text{Br}(\phi\to\nu\nu) + \text{Br}(\phi\to f) 
    \exp\left[-\left(\frac{(Q-T)^2}{m_\phi^2} - 1\right)^{-1/2}\frac{L_\text{min}}{L_\phi}\right]
    \right\}.
\end{align}
The continuous spectrum contributions in the second line above are not expected to be relevant; however, double beta decay experiments have a much stronger sensitivity to signals at the energy endpoint.

\subsubsection{Individual final state spectra and partially visible final state}

If $\phi$ decays partially visibly, i.e., to one SM photon and an invisible dark photon $\phi\to\gamma\gamma_D$, the relevant observable is the photon energy $E_3$. Likewise, certain detectors may be able to identify individual photons, electrons or positrons and infer their energies and their angles with respect to each other or to the displacement direction $\mathbf{x}_D - \mathbf{x}_P$. 

It is therefore useful to derive the spectra with respect to individual final state particle properties, which can be calculated from Eq.~\eqref{eq:distro-Egamma-costheta3-L} by integrating over $T$ (or, equivalently, $E_\phi$), and any of the other variables. Starting from the normalized distribution in $E_3$ and $\cos\theta_3$ (found by integrating Eq.~\eqref{eq:boost_relation} over $\phi_3$),
\begin{align}
&p(E_3, \cos\theta_3)dE_3 d\cos\theta_3 \nonumber \\ 
&= \frac{1}{2}\frac{|\vec{p}_3|}{\sqrt{\gamma^2(E_3 - \beta |\vec{p}_3|\cos\theta_3)^2 - m_3^2}}\delta\left(\gamma (E_3 - \beta|\vec{p}_3|\cos\theta_3) - E_3^*\right) dE_3 d\cos\theta_3 ,
\label{eq:E3_costheta3_distribution}
\end{align}
we may first integrate over $\cos\theta_3$ to find the distribution in $E_3$ only. To do so, the argument of the Dirac delta function can be rearranged for $\cos\theta_3$ as
\begin{align}
\delta\left(\gamma (E_3 - \beta|\vec{p}_3|\cos\theta_3) - E_3^*\right) = \frac{1}{\gamma\beta |\vec{p}_3|}\delta\left(\cos\theta_3 - \frac{\gamma E_3 - E_3^*}{\gamma\beta |\vec{p}_3|}\right),
\end{align}
such that the $\cos\theta_3$ integration fixes $\cos\theta_3 = (\gamma E_3 - E_3^*)/\gamma\beta |\vec{p}_3|$ and the Jacobian factor to $\mathcal{J} = |\vec{p}_3|/|\vec{p}_3^*|$. The normalized distribution in $E_3$ is then simply
\begin{align}
p(E_3)dE_3 = \frac{dE_3}{2\gamma\beta|\vec{p}_3^*|}.
\end{align}
Making explicit the maximum and minimum values of $E_3$, this may be rewritten as
\begin{align}
\label{eq:E3_pdf}
    p(E_3) dE_3 = 
    \frac{dE_3}{E_3^+(E_\phi) - E_3^-(E_\phi)} \Theta\left(E_3^-(E_\phi) < E_3 < E_3^+(E_\phi)\right)
     ,
\end{align}
with
\begin{align}
    E_3^{\pm} = \gamma(E_3^* \pm \beta |\vec{p}_3^*|) = \frac{E_\phi E_3^* \pm |\vec{p}_\phi||\vec{p}_3^*|}{m_\phi},
\end{align}
from Eq.~\eqref{eq:boost_relations}. This is the well-known flat energy distribution for a two-body decay in a boosted frame. Likewise, Eq.~\eqref{eq:E3_costheta3_distribution} may be integrated over $E_3$ by rearranging the argument of the Dirac delta function for $E_3$,
\begin{align}
\delta\left(\gamma (E_3 - \beta|\vec{p}_3|\cos\theta_3) - E_3^*\right) = \sum_{i = \pm} \frac{|\vec{p}_3^i|}{\gamma(|\vec{p}_3^i| - \beta E_3^i \cos\theta_3^i)}\delta\left(E_3 - E_3^i(\cos\theta_3)\right),
\end{align}
where the sum is over the (possibly one or two) solutions for $E_3$ as a function of $\cos\theta_3$,
\begin{align}
E_3^\pm(\cos\theta_3) = \frac{\gamma E_3^* \pm \gamma\beta \cos\theta_3 \sqrt{(E_3^*)^2 - \gamma^2m_3^2(1 - \beta^2\cos^2\theta_3)}}{\gamma^2(1 - \beta^2\cos^2 \theta_3)}.
\end{align}
For $E_\phi \geq m_\phi E_3^*/m_3$, both of the solutions are physical, while for $E_\phi < m_\phi E_3^*/m_3$, only the positive solution $E_3^+(\cos\theta_3)$ is physically relevant. Thus, we obtain the $\cos\theta_3$ distribution,
\begin{align}
p(\cos\theta_3) d\cos\theta_3 = \frac{E_3^*}{2|\vec{p}_3^*|}\frac{f(\cos\theta_3)d\cos\theta_3}{\gamma^2(1 - \beta^2\cos^2\theta_3)^2}
,
\label{eq:costheta3_distribution}
\end{align}
where
\begin{align}
f(\cos\theta_3) = \begin{cases}
\frac{\big(\beta\cos\theta_3 + \sqrt{1-\gamma^2 m_3^2(1 - \beta^2\cos^2\theta_3)/(E_3^*)^2}\big)^2}{\sqrt{1-\gamma^2 m_3^2(1 - \beta^2\cos^2\theta_3)/(E_3^*)^2}} & E_\phi < m_\phi E_3^*/m_3 \\
4\beta\cos\theta_3 & E_\phi \geq m_\phi E_3^*/m_3 \\
\end{cases}.
\end{align}
For $m_3 = 0$ (and also including the $\cos\theta_3$ limits explicitly), this yields
\begin{align}
    p(\cos\theta_3) d\cos\theta_3 = 
    \frac{1}{2} \
    \frac{d\cos\theta_3}{\gamma^2(1 - \beta\cos\theta_3)^2}\Theta\left(-1 < \cos\theta_3 < 1\right)
     .
\end{align}
With these results, we can thus determine the spectra in $E_3$ and $\cos\theta_3$ as follows.

\paragraph*{Distribution with respect to $E_3$ and $L$.} The spectrum with respect to the energy $E_3$ and displacement $L$ can be written as
\begin{align}
\label{eq:distro-Egamma-L}
    \frac{d\Gamma}{dE_3 dL} &=  
    \frac{\kappa_{0\nu}}{16\pi^2}\left(\frac{m_e}{2R}\right)^2 
    \left|\mathcal{M}_{0\nu}\right|^2 c_\nu^2 
    \frac{m_\phi^2}{|\vec{p}_3^*|L_\phi}
    \text{Br}(\phi\to f) \nonumber\\  
    &\times \int_{E_\phi^-(E_3)}^Q dE_\phi 
    \frac{g(Q - E_\phi)}{\sqrt{E_\phi^2 - m_\phi^2}} 
    \exp\left[-\left(\frac{E_\phi^2}{m_\phi^2} - 1\right)^{-1/2}\frac{L}{L_\phi}\right]
    \nonumber\\
    &\times \Theta(E_3^- < E_3 < E_3^+) \Theta(0 < L),
\end{align}
where the integration over $E_\phi$ is from the lower limit
\begin{align}
    E_\phi^-(E_3) 
    = \begin{cases}
    \frac{m_\phi}{m_3^2}\big(E_3^* E_3 - |\vec{p}_3^*|\sqrt{E_3^2 - m_3^2}\big) & m_3 \neq 0 \\
    \frac{m_\phi}{2}\Big(\frac{E_3}{E_3^*} + \frac{E_3^*}{E_3}\Big) & m_3 = 0
    \end{cases},
\end{align}
up to $E_\phi = Q$, where the minimum and maximum values of $E_3$ are given by Eq.~\eqref{eq:boost_relation-limits}. From Eq.~\eqref{eq:distro-Egamma-L}, the displacement-dependent spectrum in $E_3$ and the overall rate can be determined analogously to Eq.~\eqref{eq:distro-Tgamma} and the following expressions. If the two final state particles have the same mass, as for the $\gamma\gamma\beta\beta$ and $e^+e^-\beta\beta$ decay scenarios, the energy $E_4 = E_\phi - E_3$ is not an independent quantity. The distribution in $E_4$ is then identical to Eq.~\eqref{eq:distro-Egamma-L}.

\begin{figure}[t!]
    \centering
    \includegraphics[width=0.6\textwidth]{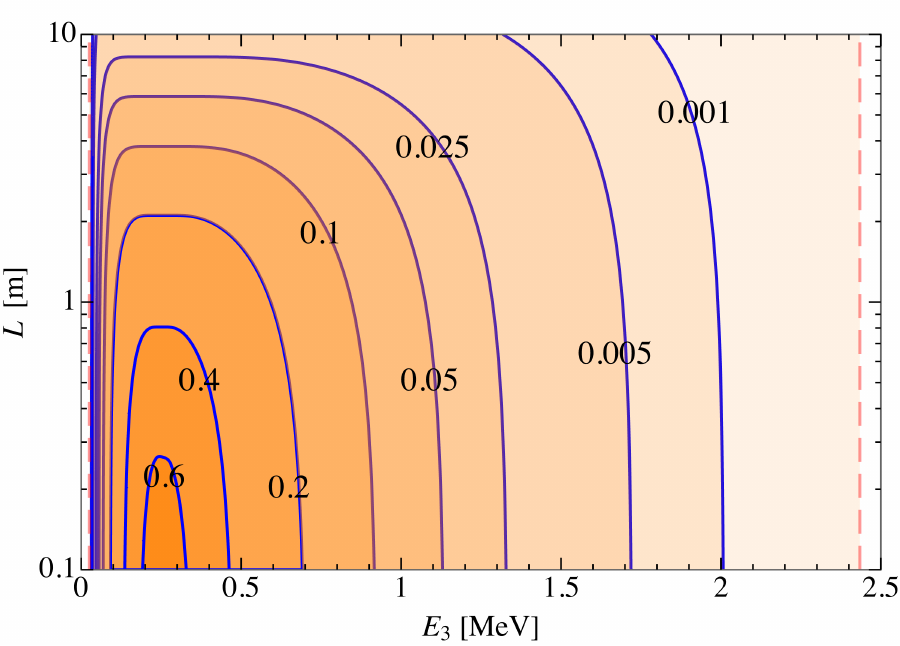}
    \caption{Normalized $\gamma\gamma'\beta\beta$ decay distribution $\Gamma^{-1}d\Gamma/(dE_3 dL)$ for $^{136}$Xe with respect to the SM photon energy $E_3$ and $\phi$ displacement $L$ with $m_\phi = 0.5$~MeV and benchmark couplings in Eq.~\eqref{eq:benchmark}.}
    \label{fig:dGdE3dL}
\end{figure}
The normalized distribution with respect to $E_3$ and $L$ is shown in Fig.~\ref{fig:dGdE3dL}, for the same benchmark $\gamma\gamma'\beta\beta$ decay scenario introduced above.
\begin{figure}[t!]
    \centering
    \includegraphics[width=0.6\textwidth]{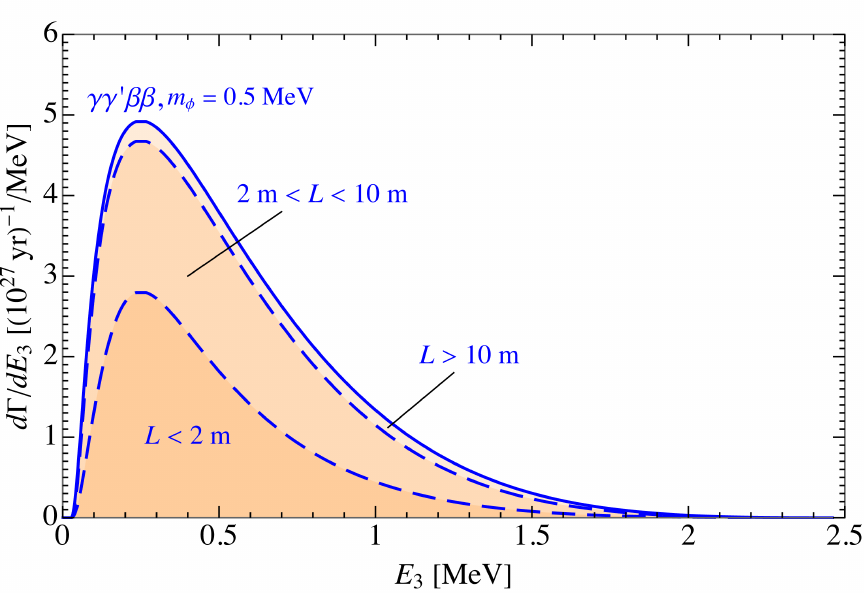}
    \caption{Distribution of $\gamma\gamma'\beta\beta$ decay for $^{136}$Xe with respect to the SM photon energy $E_3$ for $m_\phi = 0.5$~MeV and the benchmark couplings in Eq.~\eqref{eq:benchmark}. The shaded regions show the contributions where $\phi$ decays at distances $L < 2$~m, $2~\text{m} < L < 10$~m and $L > 10$~m, as indicated.}
    \label{fig:dGdE3}
\end{figure}
Likewise, the distribution with respect to $E_3$ only, but integrated over three distance ranges, is shown in Fig.~\ref{fig:dGdE3}. 

\paragraph*{Distribution with respect to photon angles and $L$.} Next, the spectrum with respect to the direction $\cos\theta_3$ and displacement $L$. For $m_3 \neq 0$, the expression using Eq.~\eqref{eq:costheta3_distribution} is too involved to present here, and thus we give the distribution in the $m_3 = 0$ scenario,
\begin{align}
\label{eq:distro-costheta3-L}
    \frac{d\Gamma}{d\cos\theta_3 dL} &=  
    \frac{\kappa_{0\nu}}{16\pi^2}\left(\frac{m_e}{2R}\right)^2 
    \left|\mathcal{M}_{0\nu}\right|^2 c_\nu^2 
    \frac{E_3^* m_\phi^3}{|\vec{p}_3^*| L_\phi} \text{Br}(\phi\to f) \nonumber\\  
    &\times \int_{m_\phi}^Q dE_\phi 
    \frac{g(Q - E_\phi)}{{\big(E_\phi - \sqrt{E_\phi^2 - m_\phi^2}\cos\theta_3\big)^2}} 
    \exp\left[-\left(\frac{E_\phi^2}{m_\phi^2} - 1\right)^{-1/2}\frac{L}{L_\phi}\right]
    \nonumber\\
    &\times \Theta(-1 < \cos\theta_3 < 1) \Theta(0 < L).
\end{align}
If the other final state particle is also massless, $m_4 = 0$, the distribution in $\cos\theta_4$ is identical to that above. 

\begin{figure}[t!]
    \centering
    \includegraphics[width=0.47\textwidth]{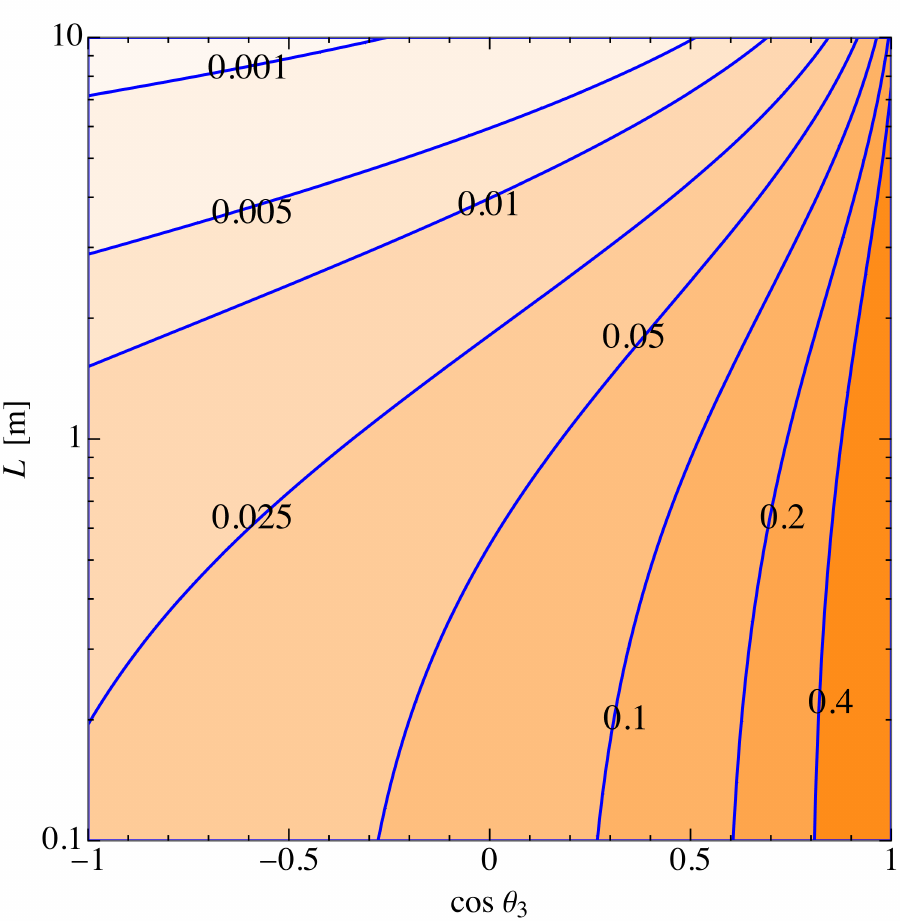}\qquad
    \includegraphics[width=0.47\textwidth]{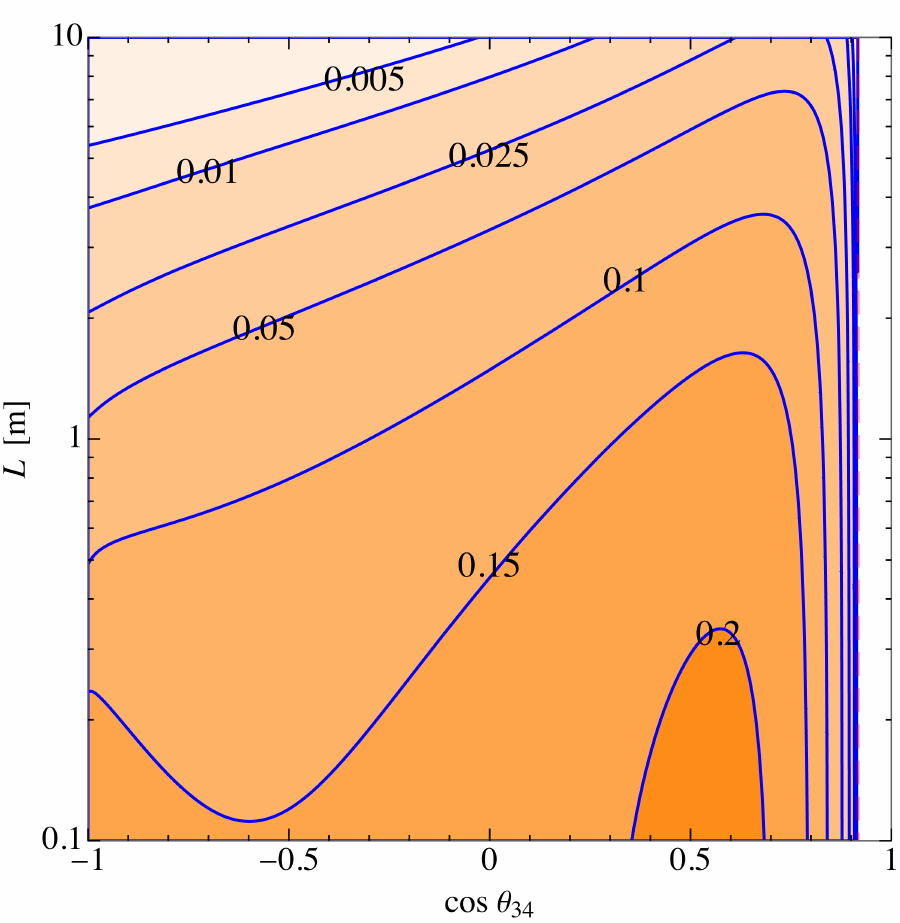}\qquad
    \includegraphics[width=0.47\textwidth]{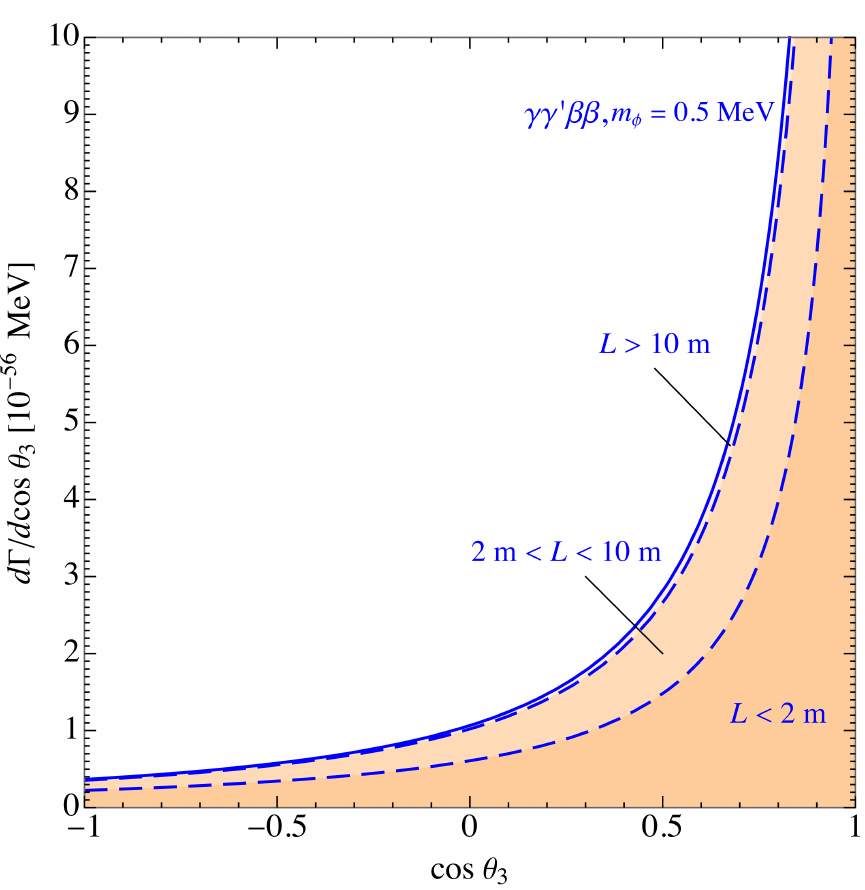}\qquad
    \includegraphics[width=0.47\textwidth]{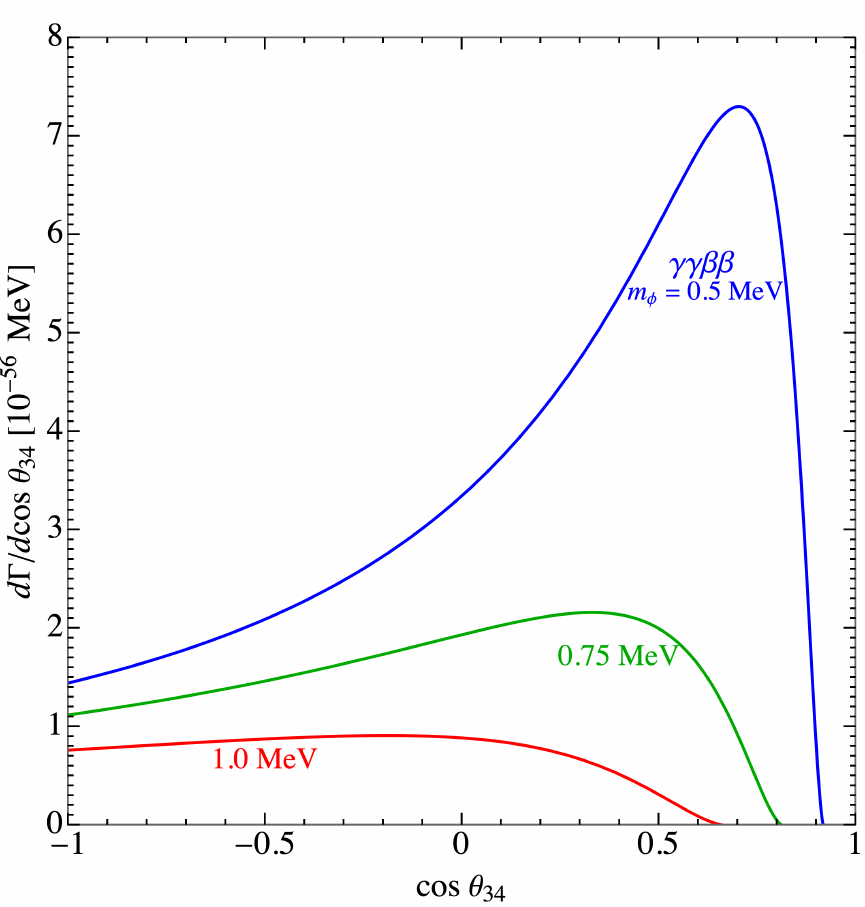}
    \caption{Normalized double differential $\gamma\gamma'\beta\beta$ spectra with respect to $L$ and the angular variables $\cos\theta_3$ (top left) and $\cos\theta_{34}$ (top right), as well as single differential distributions with respect to the same angular variables (bottom left and right, respectively) for different decay distance ranges and $\phi$ masses, as indicated. Unless specified otherwise, the pseudoscalar mass is $m_\phi = 0.5$~MeV, the dark photon mass is $m_{\gamma_D} = 0$, and the benchmark couplings in Eq.~\eqref{eq:benchmark} apply.}
    \label{fig:angular}
\end{figure}

In the scenario where both final state particles are visible, another interesting observable is the opening angle $\theta_{34}$ between the $\phi$ decay product momenta, which can be expressed in terms of the angle $\theta_3'$ in the $\phi$ rest frame as
\begin{align}
    \cos\theta_{34} = \frac{\vec{p}_3\cdot \vec{p}_4}{|\vec{p}_3||\vec{p}_4|} 
    =\frac{2E_3 E_4 + m_3^2 + m_4^2 - m_\phi^2}{2|\vec{p}_3||\vec{p}_4|}= 1 - \frac{2 m_\phi^2}{E_\phi^2 - |\vec{p}_3|^2\cos^2\theta_3'},
\end{align}
where the final equality is only valid for $m_3 = m_4 = 0$ and all other quantities are in the lab frame. The differential rate in $\cos\theta_{34}$ and the displacement $L$ for $m_3 = m_4 = 0$ is then
\begin{align}
\label{eq:distro-costheta34-L}
    \frac{d\Gamma}{d\cos\theta_{34} dL} &=  
    \frac{\kappa_{0\nu}}{8\pi^2}\left(\frac{m_e}{2R}\right)^2 
    \left|\mathcal{M}_{0\nu}\right|^2 c_\nu^2 
    \frac{m_\phi^3}{L_\phi}\frac{1}{(1-\cos\theta_{34})^{3/2}} 
    \text{Br}(\phi\to f) \nonumber\\  
    &\times \int_{E_\phi^-(c_{34})}^Q dE_\phi 
    \frac{g(Q - E_\phi)}{\sqrt{E_\phi^2 - m_\phi^2}\sqrt{E_\phi^2(1-\cos\theta_{34}) - 2m_\phi^2}} 
    \exp\left[-\left(\frac{E_\phi^2}{m_\phi^2} - 1\right)^{-1/2}\frac{L}{L_\phi}\right]
    \nonumber\\
    &\times \Theta(-1 < \cos\theta_{34} < \cos\theta_{34}^+) \Theta(0 < L).
\end{align}
where the integration over $E_\phi$ is from the lower limit
\begin{align}
    E_\phi^-(c_{34}) = \sqrt{\frac{2m_\phi^2}{1 - \cos\theta_{34}}},
\end{align}
up to $E_\phi = Q$, where $\cos\theta_{34}$ can vary from $-1$ up to the maximum value
\begin{align}
    \cos\theta_{34}^+ = 1 - \frac{2m_\phi^2}{Q^2}.
\end{align}
We note that the distribution above is only applicable for $m_3 = m_4 = 0$ and thus $\gamma\gamma\beta\beta$ decay. For massive final states ($\gamma\gamma_D\beta\beta$ and $e^+e^-\beta\beta$ decays), $\cos\theta_{34}$ is no longer a monotonic function of $\cos\theta_3'$ and it is not possible to find a closed form analytic expression as in Eq.~\eqref{eq:distro-costheta34-L}, but the distributions can be found numerically. However, as we consider an invisible dark photon in this work, $\cos\theta_{34}$ is not a measurable quantity for $\gamma\gamma_D\beta\beta$ decay.
The angular distributions described above are visualized for $\gamma\gamma'\beta\beta$ decay in Fig.~\ref{fig:angular}. 

\begin{figure}[t!]
    \centering
    \includegraphics[width=0.6\textwidth]{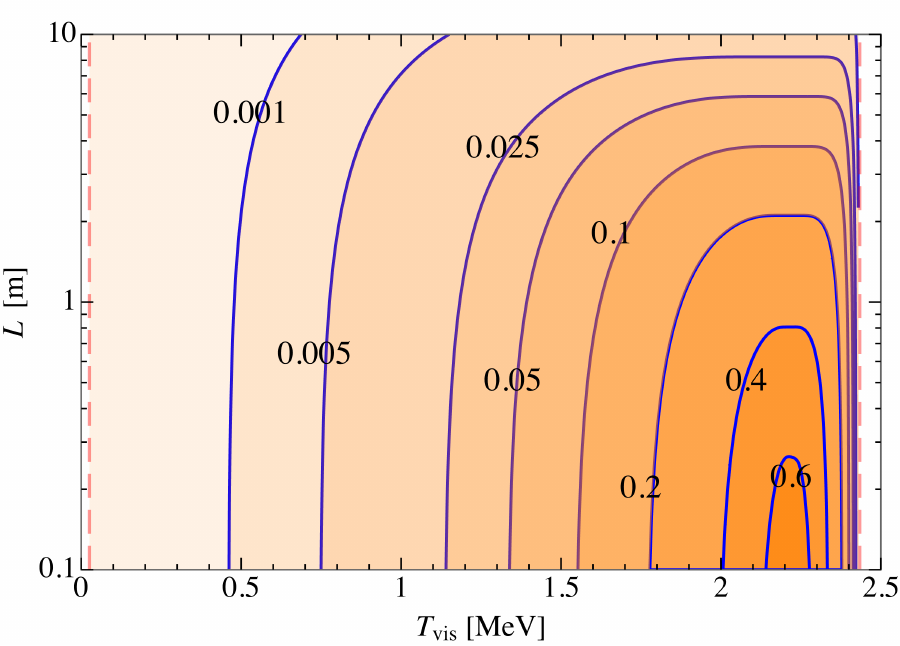}
    \caption{Normalized $\gamma\gamma_D\beta\beta$ decay distribution $\Gamma^{-1}d\Gamma/(dT_\text{vis} dL)$ in $^{136}$Xe with respect to the visible energy $T_\text{vis} = T + E_3$ and $\phi$ displacement $L$ with $m_\phi = 0.5$~MeV and benchmark couplings in Eq.~\eqref{eq:benchmark}.}
    \label{fig:dGdTvisdL}
\end{figure}

\begin{figure}[t!]
    \centering
    \includegraphics[width=0.6\textwidth]{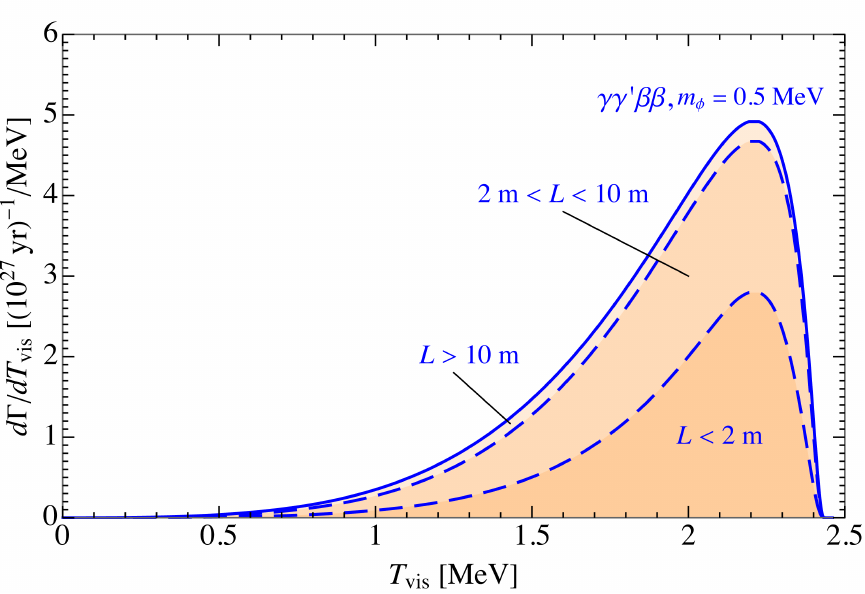}
    \caption{Distribution of $\gamma\gamma_D\beta\beta$ decay in $^{136}$Xe with respect to the visible energy $T_\text{vis} = T + E_3$ for $m_\phi = 0.5$~MeV and the benchmark couplings in Eq.~\eqref{eq:benchmark}. The shaded regions indicate the contributions where $\phi$ decays at distances $L < 2$~m, $2~\text{m} < L < 10$~m and $L > 10$~m as indicated.}
    \label{fig:dGdTvis}
\end{figure}

\paragraph*{Distribution with respect to $T_\text{vis} = T + E_3$.} In the scenario where the pseudoscalar decay is only partially visible, e.g., $\phi\to\gamma\gamma_D$, another observable of interest is the sum of electron and visible final state energies, $T_\text{vis} = T + E_3 = Q - E_4$, i.e., the total visible energy release if $\phi$ decays inside the detector. The differential rate in $T_\text{vis}$ and the displacement $L$ can be found analogously to the distribution in $E_3$ and $L$ in Eq.~\eqref{eq:distro-Egamma-L}. The probability distribution for $T_{\text{vis}}$ has the same form as Eq.~\eqref{eq:E3_pdf}, with $E_3 \to T_{\text{vis}}$ and
\begin{align}
\label{eq:boost_relation_Tvis}
    T_{\text{vis}}^{\pm} = Q - E_4^\mp = Q -  \frac{E_\phi E_4^* \mp |\vec{p}_\phi||\vec{p}_4^*|}{m_\phi}.
\end{align}
We then obtain
\begin{align}
\label{eq:distro-Tvis-L}
    \frac{d\Gamma}{dT_{\text{vis}} dL} &=  
    \frac{\kappa_{0\nu}}{16\pi^2}\left(\frac{m_e}{2R}\right)^2 
    \left|\mathcal{M}_{0\nu}\right|^2 c_\nu^2 
    \frac{m_\phi^2}{|\vec{p}_4^*| L_\phi} \text{Br}(\phi\to f) \nonumber\\  
    &\times \int_{E_\phi^-(T_\text{vis})}^Q dE_\phi 
    \frac{g(Q - E_\phi)}{\sqrt{E_\phi^2 - m_\phi^2}} 
    \exp\left[-\left(\frac{E_\phi^2}{m_\phi^2} - 1\right)^{-1/2}\frac{L}{L_\phi}\right]
    \nonumber\\
    &\times \Theta(T_\text{vis}^- < T_\text{vis} < T_\text{vis}^+) \Theta(0 < L),
\end{align}
with the integration over $E_\phi$ from the lower limit
\begin{align}
    E_\phi^-(T_\text{vis}) = 
    \begin{cases} 
    \frac{m_\phi}{m_4^2}\Big[ E_4^*(Q - T_{\text{vis}}) - |\vec{p}_4^*|\sqrt{(Q - T_{\text{vis}})^2 - m_4^2}\Big] & m_4 \neq 0 \\
    \frac{m_\phi}{2}\Big(\frac{Q-T_{\text{vis}}}{E_4^*} + \frac{E_4^*}{Q - T_{\text{vis}}}\Big) & m_4 = 0
    \end{cases},
\end{align}
up to $E_\phi = Q$, which can be substituted into Eq.~\eqref{eq:boost_relation_Tvis} to obtain the minimum and maximum possible values of $T_{\text{vis}}$. The distribution in Eq.~\eqref{eq:distro-Tvis-L} can be integrated over $L$ to obtain the spectrum in the range $L_1 < L < L_2$ and further over $T_{\text{vis}}$ to obtain the total decay rate within a certain distance interval, $\Gamma(L_1 < L < L_2)$. The spectrum in the total energy $T_{\text{tot}}$ deposited in a finite size detector can be constructed analogously to Eq.~\eqref{eq:distro_Ttotal_full} as,
\begin{align}
\label{eq:distro_Ttotal_partial}
    \left.\frac{d\Gamma}{dT_\text{tot}}\right|_0^{L_\text{min}} &=
    \left.\frac{d\Gamma}{dT_{\text{vis}}}\right|_{0}^{L_\text{min}}  \nonumber\\
    &\hspace{-1em}+ \frac{d\Gamma_\phi}{dT} \Bigg\{
    \text{Br}(\phi\to\nu\nu) + \text{Br}(\phi\to f)
    \exp\left[-\left(\frac{(Q-T)^2}{m_\phi^2} - 1\right)^{-1/2}\frac{L_\text{min}}{L_\phi}\right]
    \Bigg\}.
\end{align}
The spectra for the $\gamma\gamma_D\beta\beta$ decay scenario are shown in Figs.~\ref{fig:dGdTvisdL} and \ref{fig:dGdTvis}. While the visible energy release will not be at the double beta decay endpoint, the spectrum $d\Gamma/dT_\text{vis}$ peaks at a significantly higher energy than other exotic decay spectra described in the literature. In the example shown, with $m_\phi = 0.5$~MeV and $m_{\gamma_D} = 0$, the peak is at $T_\text{vis} \approx 2.2$~MeV compared to $T_\text{vis} \approx 1.9$~MeV for standard (massless) Majoron emission, see Fig.~\ref{fig:dGdT}.

\section{Statistical procedure and experimental setups}

Our goal is to estimate the sensitivity of current and future double beta decay experiments to the combined signal arising from the bulk, prompt, and displaced decay of the pseudoscalar $\phi$ into different final states ($\gamma\gamma$, $\gamma\gamma_D$ and $e^+e^-$). We follow the standard frequentist approach~\cite{Tanabashi:2018oca} in this work and employ a log-likelihood analysis. In contrast to the analysis of Ref.~\cite{Boudjema:2025okq}, where the BSM contribution was contained entirely in the continuous electron energy spectrum, we now distinguish three experimentally different contributions depending on the displacement of $\phi$ inside the detector.

We first consider the bulk contribution, corresponding to the standard pseudoscalar emission with a continuous electron energy spectrum, as discussed in Ref.~\cite{Boudjema:2025okq}. We denote the resulting log-likelihood by $-2\ln \mathcal{L}_\text{bulk}(c_\nu)$ where this function depends on a single parameter $c_\nu$ for a fixed pseudoscalar mass $m_\phi$. This analysis neglects the bulk decays to semi-invisible decays to $\gamma \gamma^\prime$, but should be a very good approximation if $\mathrm{Br}(\phi\to\nu\nu)$ becomes the dominant decay channel of the pseudoscalar. Even when the branching fractions into neutrinos and photons are comparable, the resulting modification of the electron energy spectrum is expected to be small, and the approximation remains adequate for our sensitivity estimation. The relevant data to estimate the log-likelihood for the bulk contribution and for the considered isotope $^{136}$Xe have been provided in Tab.~\ref{tab:isotopes}, where $m_{\text{iso}}$ and $(T^{2\nu}_{1/2})_{\text{exp}}$ correspond to the molar mass and experimental $2\nu\beta\beta$ decay half-life. Furthermore, as we treat the $2\nu\beta\beta$ decay background and the relevant signal $0\nu\beta\beta$ decay NMEs as individual random variables, $\langle \mathcal{M}_{2\nu}\rangle$ and $\langle \mathcal{M}_{0\nu}\rangle$ denote the average best-fit values for the corresponding NMEs, respectively, while the theoretical uncertainties are $\delta \mathcal{M}_{2\nu}$ and $\delta \mathcal{M}_{0\nu}$.

\begin{table}[t!]
	\setlength{\tabcolsep}{6pt}
	\begin{tabular}{ccc|cccc}
		\hline
		 $m_\text{iso}$~[g/mol] & $Q$~[MeV] & $(T_{1/2}^{2\nu})_\text{exp}$~[yr] & $\langle \mathcal{M}_{2\nu}\rangle$ & $\delta\mathcal{M}_{2\nu}$ & $\langle \mathcal{M}_{0\nu}\rangle$ & $\delta\mathcal{M}_{0\nu}$ \\\hline
		 131 & 2.46 & $2.2\times 10^{21}$ & 0.016 & 0.19 & 2.6 & 1.1 \\\hline
	\end{tabular}
	\caption{Molar mass $m_\text{iso}$, $Q$ value and experimental $2\nu\beta\beta$ decay half-life $(T_{1/2}^{2\nu})_\text{exp}$ for the $^{136}$Xe isotope considered in this work \cite{Boudjema:2025okq}. Also given are the average NMEs $\langle \mathcal{M}_{2\nu}\rangle$, $\langle \mathcal{M}_{0\nu}\rangle$ and the standard deviations $\delta\mathcal{M}_{2\nu}$, $\delta\mathcal{M}_{0\nu}$.}
	\label{tab:isotopes}
\end{table}

In addition to the bulk contribution, the pseudoscalar $\phi$ can decay into different completely visible or semi-invisible final states within the fiducial volume of the detector, giving rise to a displaced signal. We treat this contribution as a single counting bin, defined by $\phi$ decays occurring between the minimum resolvable displacement $L_\text{min}$ for the displaced vertex signature and the effective detector size $L_\text{max}$. The corresponding partial decay rate is denoted by
$\Gamma_\text{displ} \equiv \Gamma(L_\text{min}<L<L_\text{max})$ from Eq.~\eqref{eq:totGamma},
where the dependence on $L_\text{min}$ and $L_\text{max}$ includes the appropriate decay probability and boost distribution of the pseudoscalar. The expected number of displaced events is therefore
\begin{align}
N_\text{displ} = N_A\cdot\mathcal{E} \cdot m_\text{iso}^{-1} \cdot \Gamma_\text{displ}.
\end{align}
Here, $N_A = 6.022 \times 10^{23}$ mol$^{-1}$ is Avogadro's number, and $\mathcal{E}$ denotes the fiducial exposure of the considered experiment in units of kg$\cdot$yr. We can safely consider that the displaced signal is effectively background-free and use a Poisson distribution to estimate the corresponding log-likelihood for a displaced signature,
\begin{align}
-2\ln\mathcal{L}_{\text{displ}} (c_\nu, g_{\phi\gamma\gamma^\prime}) = 2N_{\text{displ}}.
\end{align}
This is based on the general Poisson probability distribution for the observed number of events $n$ given the expected number of events $\lambda = \lambda_\text{sig} + \lambda_\text{bkg}$,
\begin{align}
        \mathcal{L}(n) = \frac{\lambda^n e^{-\lambda}}{\Gamma(n+1)},
\end{align}
with $\lambda_\text{bkg} = n = 0$ (background free and assuming no events are observed as we want to get the exclusion sensitivity). Then, $-2\ln \mathcal{L}(0) = 2\lambda_\text{sig} = 2N_{\text{displ}}$.

Finally, pseudoscalar decays occurring before the minimum resolvable displacement $L_{\text{min}}$ contribute to a prompt signal, only possible for completely visible final states with a photon or $e^+e^-$ pair. We treat these events as a single counting bin localized at the endpoint of the electron energy spectrum. The corresponding decay rate is $\Gamma_\text{prompt} \equiv \Gamma_{\gamma\gamma}(0<L<L_\text{min})$,
and the expected number of prompt events can be expressed as,
\begin{align}
N_\text{prompt} = N_A \cdot \mathcal{E} \cdot m_\text{iso}^{-1} \cdot
\Gamma_\text{prompt}.
\end{align}
In contrast to the displaced search, the prompt contribution is accompanied by a non-zero background. Following the treatment of Ref.~\cite{Agostini:2022bjh}, we characterize this background by a rate $\mathcal{B}$, such that
\begin{align}
N_\text{bkg}=\mathcal{E} \cdot \mathcal{B}.
\end{align}
Assuming that the observed number of events is given by the expected background, $n=N_\text{bkg}$, the Poisson likelihood for the signal-plus-background hypothesis gives
\begin{align}
-2 \ln\mathcal{L}_\text{prompt} (c_\nu, g_{\phi\gamma\gamma^\prime})
& = 2\left[N_\text{prompt} + N_\text{bkg} \ln\left(\frac{N_\text{bkg}}{N_\text{bkg} + N_\text{prompt}}\right)\right].
\end{align}
This corresponds to an Asimov data set in which the observed prompt spectrum agrees with the expected background-only prediction.

The three contributions correspond to mutually exclusive classes of pseudoscalar decay signatures, distinguished by their decay position. We therefore combine them at the level of the likelihood, or equivalently at the level of the corresponding log-likelihood ratios. Since the bulk contribution depends only on the pseudoscalar-neutrino coupling $c_\nu$, whereas the prompt and displaced contributions additionally depend on the radiative coupling $g_{\phi\gamma\gamma^\prime}$, the effective dimensionality of the parameter space changes depending on the size of the latter coupling. In particular, in the limit of sufficiently small $g_{\phi\gamma\gamma^\prime}$, the pseudoscalar predominantly contributes through the bulk channel and the sensitivity is effectively determined by a single parameter, $c_\nu$. In this regime, the appropriate 90\%~CL threshold for the profile likelihood is $-2\ln \mathcal{L}=2.71$. When the radiative decay becomes relevant, both $c_\nu$ and $g_{\phi\gamma\gamma^\prime}$ contribute to the observable signal, and the corresponding 90\%~CL threshold for two parameters is $-2\ln \mathcal{L} = 4.61$. To account for this change in the effective number of degrees of freedom, we define the combined test statistic as
\begin{align}
\label{eq:chi2_combined}
-2\ln \mathcal{L}(c_\nu,g_{\phi\gamma\gamma^\prime}) = -2\left[\frac{\ln \mathcal{L}_\text{bulk}(c_\nu)}
{n_{\text{th}}^{\text{bulk}}} + \frac{\ln \mathcal{L}_\text{displ} (c_\nu,g_{\phi\gamma\gamma^\prime})+\ln \mathcal{L}_\text{prompt}(c_\nu,g_{\phi\gamma\gamma^\prime})}{n_{\text{th}}^{\text{non-bulk}}}\right],
\end{align}
where,
\begin{align}
n_{\text{th}}^{\text{bulk}}=2.71,
\qquad
n_{\text{th}}^{\text{non-bulk}}=4.61.
\end{align}
Here, $n_{\text{th}}^{\text{bulk}}$ corresponds to the 90\%~CL threshold for one degree of freedom, while $n_{\text{th}}^{\text{non-bulk}}$ corresponds to the corresponding threshold for two degrees of freedom. With this normalization, the exclusion condition can be expressed in terms of a common threshold,
\begin{align}
\label{eq:exclusion_condition}
-2\ln \mathcal{L}(c_\nu,g_{\phi\gamma\gamma^\prime})=1.
\end{align}
In the limit $g_{\phi\gamma\gamma^\prime}\to0$, the prompt and displaced contributions vanish and Eq.~\eqref{eq:chi2_combined} reduces to
\begin{align}
-2\ln \mathcal{L}(c_\nu,0) = \frac{-2\ln \mathcal{L}_\text{bulk}(c_\nu)}{2.71},
\end{align}
so that the 90\%~CL exclusion condition $-2\ln \mathcal{L}(c_\nu,g_{\phi\gamma\gamma^\prime})=1$ is equivalent to $-2\ln \mathcal{L}_\text{bulk}=2.71$, as required for a one-parameter limit. Conversely, when the non-bulk contribution is relevant, the corresponding two-parameter criterion is governed by the 90\%~CL threshold $-2\ln \mathcal{L}(c_\nu,g_{\phi\gamma\gamma^\prime})=4.61$. This prescription therefore allows the sensitivity contours to interpolate between the one-parameter bulk-dominated regime and the two-parameter regime in which the prompt and displaced channels provide additional sensitivity to the radiative coupling.

\begin{table}[t!]
	\setlength{\tabcolsep}{3pt}
	\begin{tabular}{cccccccc}
		\hline
		Experiment & $\mathcal{E}$~[$\text{kg}\cdot\text{yr}$] & $T_\text{min}$~[keV] &
        $\Delta T$~[keV] & $\sigma_f$~[\%] &
        $L_\text{min}$~[m] &
        $L_\text{max}$~[m] &
        $\mathcal{B} [(\text{kg}\cdot\text{yr})^{-1}]$
        \\\hline
		 KamLAND-Zen~\cite{KamLAND-Zen:2019imh} & $126$ & 100 & $50$ & $0.3$ & $0.1$ & $4.0$ & $10^{-4}$ \\
		 nEXO~\cite{nEXO:2021ujk} & $1.9\times 10^4$ & 100 & $5$ & $0.5$ & 0.1 & 18.6 & $10^{-4}$ \\
         n$^2$EXO & $1.9\times 10^6$ & 100 & $5$ & $0.5$ & 0.1 & 86.3 & $10^{-4}$ \\
        \hline
	\end{tabular}
	\caption{Current or expected future exposure $\mathcal{E}$, minimum electron energy $T_\text{min}$, energy resolution (bin width) $\Delta T$, relative systematic uncertainty $\sigma_f$, and background rate $\mathcal{B}$ of the $^{136}$Xe double beta decay experiments considered in this work. Here, $L_{\rm{min}}$ ($L_{\rm{max}}$) corresponds to the minimum (maximum) distance needed to observe a displaced signature.}
	\label{tab:setups}
\end{table}

Now, to employ this statistical approach to put 90\%~CL contours in the parameter spaces spanned by different relevant couplings for $\gamma\gamma\beta\beta, \gamma\gamma_D\beta\beta$ and $e^+e^-\beta\beta$ decay processes, we have considered three different experimental setups using the $^{136}$Xe isotope: KamLAND-Zen, with a current $0\nu\beta\beta$ decay exclusion of $T_{1/2}^{0\nu} > 10^{26}$ yr; nEXO, with projected sensitivity of $T_{1/2}^{0\nu} > 10^{28}$ yr; and n$^2$EXO, a hypothetical far-future setup with projected sensitivity $T_{1/2}^{0\nu} > 10^{30}$ yr. The relevant experimental details used in our analysis have been presented in Tab.~\ref{tab:setups}. Here, we have $T_{\text{min}}$ and $\Delta T$, which denote the minimum electron energy and energy resolution or bin width, respectively. Moreover, $\sigma_f$ is the relative systematic experimental uncertainty, which we have taken to be constant for a given experiment.

\bibliography{references}
\end{document}